\documentclass[%
 reprint,
 superscriptaddress,
 amsmath,amssymb,
 aps,
]{revtex4-2}
\usepackage{graphicx}% Include figure files
\usepackage{dcolumn}% Align table columns on decimal point
\usepackage{bm}% bold math
\usepackage{accents}
\usepackage{hyperref}% add hypertext capabilities
 \usepackage{tikz}
 \usetikzlibrary{positioning}
  \usetikzlibrary{calc}
\usepackage{physics}% bra, ket
\usepackage{xcolor}
\usepackage{mathtools}
\usepackage{amssymb,amsmath}
\usepackage{color,soul}
\usepackage{comment}

\usepackage{subcaption}
\usepackage{caption}

\DeclareMathOperator*{\argmax}{arg\,max}

\usepackage[makeroom]{cancel}
\usepackage{dsfont}
\usepackage[normalem]{ulem}
\usepackage{amsmath,amsfonts}

\usepackage{ragged2e}

\newenvironment{proof-of}[1]{\medskip\noindent\textbf{Proof of {#1}.}}{\hfill$\blacksquare$\medskip}

\begin{document}

\title
{Quantum thermometry with $k$-local Hamiltonians}


%\author{Ayaka Usui}
%\email{ayaka.usui@uab.cat}
% \affiliation{Departament de F\'{i}sica Qu\`{a}ntica i Astrof\'{i}sica, Institut de Ci\`{e}ncies del Cosmos (ICCUB), Facultat de F\'{i}sica, Universitat de Barcelona, Mart\'{i} i Franqu\'{e}s, 1, E08028 Barcelona, Spain}
 %\affiliation{Grup d'\`{O}ptica, Departament de F\'{i}sica, Universitat Aut\`{o}noma de Barcelona, 08193 Bellaterra, Spain}

 \author{Jorge López González}
\email{jorge.lgonzalez@alumnos.upm.es}
 \affiliation{Quantum Mathematics, Algorithms and Machine Intelligence (QUMAMI), Departamento de Matem\'atica Aplicada a la Ingenier\'ia Industrial,
Universidad Polit\'ecnica de Madrid, 28006 Madrid, Spain}

 \author{Mohammad Mehboudi}
\email{mohammad.mehboudi@tuwien.ac.at}
 \affiliation{Technische Universität Wien, Atominstitut, Vienna Center for
Quantum Science and Technology, Stadionallee 2, 1020 Vienna, Austria}

 \author{Anna Sanpera}
 \email{anna.sanpera@uab.cat}
 \affiliation{F\'{\i}sica Te\`{o}rica: Informaci\'{o} i Fen\`{o}mens Qu\`{a}ntics, 
	Departament de F\'{\i}sica, Universitat Aut\`{o}noma de Barcelona, 08193 Bellaterra, Spain}
\affiliation{ICREA, Pg. Llu\'is Companys 23, 08010 Barcelona, Spain}

 \author{Mar\'ia Garc\'ia D\'iaz}
\email{maria.garcia.diaz@upm.es}
 \affiliation{Quantum Mathematics, Algorithms and Machine Intelligence (QUMAMI), Departamento de Matem\'atica Aplicada a la Ingenier\'ia Industrial,
Universidad Polit\'ecnica de Madrid, 28006 Madrid, Spain}

\date{\today}

\begin{abstract}
In equilibrium quantum thermometry, the precision of temperature estimation increases with the heat capacity of the probe thermometer.
For an $n$-body thermometer, the maximum heat capacity is attained by a Hamiltonian that requires $n$-local interactions, making its physical implementation challenging. In the thermodynamic limit, the optimal heat capacity scales quadratically with $n$. Although more experimentally accessible 2-local Hamiltonians can achieve this optimal scaling, their performance remains significantly below the theoretical maximum for realistic system sizes. 
Here, we restrict to this scenario and show a substantial enhancement in  the thermometric performance  with only a minimal increase in locality in some Hamiltonian models. 
We further introduce \mbox{$k$-local} generalizations of these models and show that their thermometric performance systematically approaches the theoretical optimum as the locality is increased. By analyzing their spectra, we relate this enhancement to the progressive emergence of the spectral structure underlying optimal equilibrium thermometry. Lastly, we discuss the heat capacity as a witness of Hamiltonian locality.
\end{abstract}

\maketitle

\section{Introduction}\label{sec:intro}
Advances in the control of quantum systems have enabled temperature measurements at increasingly small length scales and in regimes where conventional thermometry becomes challenging~\cite{Giazotto2006,Yue2012,Mehboudi2019,DePasquale2018}. A variety of quantum platforms are now being explored for this purpose, ranging from nanoscale thermometers for biological environments~\cite{Kucsko2013,Fujiwara2020}, and nanoscale electron calorimeters sensitive to individual energy quanta~\cite{Gasparinetti2015,Halbertal2016,Karimi2020} to single-atom thermometric probes~\cite{Hohmann2016,Bouton2020,Adam2022}. In parallel, considerable theoretical effort has been devoted to establishing the fundamental limits of temperature estimation and identifying quantum probes capable of approaching them, both at and away from thermal equilibrium~\cite{Hovhannisyan2018,Potts2019,Jorgensen2020,Mukherjee2019, Glatthard2022,Correa2017,Mehboudi2019Polarons,Planella2022,Khan2022,Brunelli2011,Brunelli2012,Jevtic2015,Guo2015, DePasquale2017,Hofer2017,Cavina2018,Mancino2020, Mitchison2020,Hovhannisyan2021,Zhang2022}. In particular, the spectral structure of optimal thermometers has been identified~\cite{Correa2015,Campbell2018,Mok2021,Sekatski2022}, revealing a quadratic scaling of thermometric precision with the number of constituents~\cite{Mehboudi2022}.

Here, we focus on the broadly applicable setting of equilibrium thermometry~\cite{Mehboudi2019}, where the accuracy of temperature estimation is governed by the heat capacity of the probe. Maximizing the heat capacity therefore yields the highest thermometric precision. However, the corresponding optimal spectrum for an $n$-constituent probe requires $n$-body interactions \cite{Correa2015,Mok2021}, making its physical realization challenging. This limitation motivates the search for physically realizable probes that can approach optimal thermometric performance using interactions of limited complexity. Previous work has shown that suitably engineered 2-local Hamiltonians can recover the optimal quadratic scaling of the heat capacity in the thermodynamic limit~\cite{Abiuso_2024}. Nevertheless, for small system sizes, their performance remains significantly below the theoretical optimum. This naturally raises the question of whether a moderate increase in interaction locality can improve finite-size performance while preserving the optimal asymptotic scaling.

In this work, we address this question by exploring thermometric probes described by $k$-local Hamiltonians. We first focus on the transition from two- to three-local interactions, identifying configurations that substantially enhance the heat capacity at finite size, while  retaining the optimal quadratic scaling in the thermodynamic limit. We then extend these constructions to arbitrary locality $k$ and study how their thermometric performance evolves as the locality is increased. Our analysis shows that the heat capacity progressively approaches the theoretical optimum with increasing $k$. By examining the corresponding energy spectra, we connect this behavior to their gradual convergence toward the spectral structure of the optimal thermometer.

The rest of the paper is organized as follows. In Sec.~\ref{preliminaries}, we introduce the theoretical framework. We first review $k$-local Hamiltonians, and summarize the principles of equilibrium quantum thermometry. Finally, we discuss previous results on optimal $n$-qubit thermometers, considering both unrestricted and two-local Hamiltonians. In Sec.~\ref{3local}, we investigate the optimization of the heat capacity under three-local interactions and identify Hamiltonian structures that enhance thermometric performance. In Sec.~\ref{klocal}, we generalize these constructions to arbitrary $k$-local interactions and characterize their spectra. We conclude in Sec.~\ref{discussion} with a discussion of our results.

\section{Preliminary concepts}\label{preliminaries}
\subsection{$k$-local Hamiltonians}

Few-body interactions, and two-body  in particular, constitute the basic building blocks of many quantum many-body Hamiltonians. Yet, important classes of systems feature higher-order interactions. Prominent examples include lattice gauge theories and topological models such as Kitaev’s toric code, whose Hamiltonians naturally contain multibody interaction terms. To fix ideas, we denote by ${\cal{H}}_N$  an $N$-dimensional Hilbert space and by $\mathcal{B}({\cal{H}}_N)$  the set of  its bounded operators. An $n$-qudit \emph{$k$-local} Hamiltonian $H\in \mathcal{B}({\cal{H}}_{d^n})$ is of the form \mbox{$H=\sum_{i=1}^{M}h_i$}, where $h_i$ is a Hamiltonian acting non-trivially on at most  $k$ qudits, and $M$ is some positive integer. Note that  $k$-locality does not
necessarily imply
interaction among $k$ next neighbouring objects, and that there is no restriction on spatial locality. Note also that, for $k^{\prime}>k$, any $k$-local Hamiltonian is also $k^{\prime}$-local. We then say that all $k$-local Hamiltonians are \textit{more local} than any $k^{\prime}$-local Hamiltonian that is not $k$-local. This induces a hierarchy among the set of Hamiltonians based on their locality. To make this concrete, let us restrict to local dimension $d=2$ (qubits). Then, define
the manifold of $n$-qubit $k$-local Hamiltonians as
\begin{equation}
\boldsymbol{H}_{n,k} \coloneqq \big\{H\in \mathcal{B}({\cal{H}}_{2^n}) |~ H=\sum_{i=1}^{M^{(k)}}\alpha_i \Lambda_i, ~\Lambda_i \in {\bm \Lambda}^{(k)}\big\}, 
\end{equation}
where ${\bm \Lambda}^{(k)} = {\bigcup}_{j=0}^k\{G_i^{(j)}\}_{i=1}^{L^{(j)}}$ is the union of the sets of all possible \textit{genuinely} $j$-body generators---that is, tensor products of exactly $j$ Pauli matrices ($X, Y, Z$) and $n-j$ identity matrices, also referred to as a Pauli string with \textit{weight} $k$. Note that $L^{(j)}=3^j\binom{n}{j}$, since there are $\binom{n}{j}$ ways to choose
the $j$ qubits and three Pauli matrices on each, and $M^{(k)}=\sum_{j=0}^{k}L^{(j)}$.
In particular, $M^{(n)}=4^n=2^{2n}$ is the dimension of the real vector space of
Hermitian operators on $n$ qubits, so that $\boldsymbol{H}_{n,n}$ contains all $n$-qubit
Hamiltonians. Then, $\bm{H}_{n,k}\subset\bm{H}_{n,k'}$ for $k<k'\le n$. Lastly, Hermiticity of the Hamiltonian requires that the coefficients $\{\alpha_i\}_{i=1}^{M^{(k)}}$ are real.
Within each manifold of Hamiltonians, additional constraints on $\{\alpha_i\}_{i=1}^{M^{(k)}}$  define different classes of Hamiltonians. We denote by $\boldsymbol{H}_{n,k}^\chi$ the set of $n$-qubit $k$-local Hamiltonians of class $\chi$, and by $H_{n,k}^\chi$ a Hamiltonian in this set. 

\subsection{Equilibrium thermometry}

We consider the estimation of the temperature $T$ of a sample using a quantum system as a thermometer. The thermometer is brought into thermal contact with the sample and allowed to equilibrate. Assuming that the equilibrium state of the thermometer is described by the canonical ensemble, its state is the Gibbs state
\begin{align} 
\rho_\beta(H) & = \frac{e^{-\beta H}}{\tr(e^{-\beta H})} = \sum_i p_i |E_i\rangle\langle E_i|, \nonumber\\
\text{with~} p_i & = \frac{e^{-\beta E_i}} {\sum_j e^{-\beta E_j}}. \label{eq:gibbs_state} 
\end{align} 
where $H=\sum_i E_i |E_i\rangle\langle E_i|$ is the spectral decomposition of the Hamiltonian of the thermometer and \mbox{$\beta=(k_{\mathrm B}T)^{-1}$} is the inverse temperature. Throughout this work, we set $k_{\mathrm B}=1$. 
The temperature is encoded in the populations $p_i$ while the basis $\{\ket{E_i}\}_i$ itself is temperature independent. Therefore, the temperature can be optimally inferred from energy measurements on the thermalized probe. After repeating the energy measurements $\nu$ times (iid) one observes outcomes ${\bm x} = \{x_\nu,\dots,x_1\}$ with $x_k$ labeling the energy level observed in the $k$th measurement. One then builds an \textit{estimator} function ${\tilde T}({\bm x})$ that maps measurement outcomes to temperature values. Inevitably, due to the statistical uncertainty in measurement outcomes, the estimated temperature will be imperfect. In the frequentist approach to thermometry, the mean square error (MSE) quantifies the precision of the estimator
\begin{align}
    (\Delta {\tilde T})^2 \coloneqq  \sum_{\bm x}p({\bm x})({\tilde T}({\bm x}) - T)^2,
\end{align}
with $p({\bm x})$ being the likelihood of obtaining outcome ${\bm x}$. For iid measurements, the likelihood is simply $p({\bm x}) = \prod_{k=1}^{\nu} p_{x_k}$, with $p_{x_k}$ described according to the Gibbs distribution Eq.~\eqref{eq:gibbs_state}. The Cram\'er-Rao bound ~\cite{Cramer1999,BraunsteinCaves1994} states that for unbiased estimators---those satisfying $\sum_{\bm x}p({\bm x}) {\tilde T}({\bm x}) = T$---the MSE is lower bounded by the inverse of the Fisher information
\begin{align}
    (\Delta {\tilde T})^2 \geq \frac{1}{\nu {\cal F}_T},
\end{align}
with ${\cal F}_T$ being the [per measurement] quantum Fisher information (QFI).
For a Gibbs state, the QFI~\cite{Mehboudi2019,Correa2015} associated with $T$ is given by
\begin{equation} 
\mathcal{F}_T = \frac{C(H,\beta)}{T^2}, \label{eq:qfi_temperature} 
\end{equation} 
where $C(H,\beta)$ denotes the heat capacity of the thermometer which we will discuss below. The Cram\'er--Rao bound can be rewritten as \begin{equation} 
\frac{(\Delta {\tilde T})^2}{T^2} \geq \frac{1}{\nu C(H,\beta)}. \label{eq:crb_temperature} 
\end{equation}
This inequality establishes a direct connection between thermometric precision of a probe to its heat capacity: at fixed temperature, a larger heat capacity corresponds to a larger QFI and hence to a smaller lower bound on the error-to-signal ratio. This lower bound can be saturated in the asymptotic limit of large data ($\nu \gg 1$)---which is our focus---by using the maximum likelihood estimator. As such, throughout we will take the quantum Fisher information or, equivalently, the heat capacity of the probe as our figure of merit for precision, and we aim to maximise it in order to achieve the minimum error-to-signal ratio.

The heat capacity is defined as the temperature derivative of the mean energy, \begin{equation} 
C(H,\beta) \coloneqq \frac{d}{dT}\langle H\rangle_\beta = -\beta^2 \frac{d}{d\beta} \langle H\rangle_\beta,
\label{eq:heat_capacity_definition} 
\end{equation} 
with $\left\langle O \right\rangle_{\beta}\coloneqq {\rm tr}[O \rho_{\beta}(H)]$ for any operator $O$. A direct evaluation of Eq.~\eqref{eq:heat_capacity_definition} then yields 
\begin{equation} 
C(H,\beta) =  \beta^2 (\Delta_{\beta} H)^2= \beta^2\left[\sum_i p_i E_i^2 - \Big[ \sum_i p_i E_i\Big]^2 \right],\label{eq:heat_capacity_variance} \end{equation} 
where for any operator $O$ we define $(\Delta_{\beta} O)^2 \coloneqq \left\langle (O - \langle O\rangle_{\beta})^2\right\rangle_{\beta}$.
This expression shows that the heat capacity is fully determined by the spectrum of the Hamiltonian and the inverse temperature. It is useful to absorb the temperature scale into the Hamiltonian. In particular, for any $\gamma>0$, $C(\gamma H,\gamma^{-1}\beta)=C(H,\beta)$. We therefore introduce the dimensionless Hamiltonian 
$\widetilde H\coloneqq \beta H$, such that
$C(H,\beta) = C(\widetilde H,1)$. Hence, at a fixed target temperature, the optimization problem can be expressed entirely in terms of the dimensionless Hamiltonian $\widetilde H$. In the following, we drop the tilde for notational simplicity and write
$C(H) \coloneqq  C(H,1)$ and $(\Delta H)^2\coloneqq (\Delta_1 H)^2$, so that  $C(H)=(\Delta H)^2$. 
\subsection{Optimal $n$-qubit thermometers: from $n$-local to 2-local Hamiltonians}

As a consequence of Eq.~(\ref{eq:crb_temperature}), the optimization of an equilibrium thermometer can be formulated as the maximization of its heat capacity at the target temperature. Assuming access to arbitrary $k$-local Hamiltonians, finding the optimal thermometer thus amounts to solving the following maximization:
\begin{equation}
    \accentset{\ast}{H}_{n,k} \coloneqq \argmax_{H\in \boldsymbol{H}_{n,k} }(\Delta H)^2.
\end{equation} 
This can be  generalized to scenarios with additional constraints on top of locality:
\begin{equation}
    \accentset{\ast}{H}_{n,k}^{\chi} \coloneqq \argmax_{H\in \boldsymbol{H}_{n,k}^{\chi} }(\Delta H)^2.
\end{equation} 
For $n$-local Hamiltonians, the solution  was analytically derived in~\cite{Correa2015} and has the following spectrum:
\begin{equation}
\accentset{\ast}{H}_{n,n} = 0\, \ketbra{0}{0} +  \sum_{i=1}^{2^n-1} x \ketbra{i}{i},
\label{eq: H_opt}
\end{equation}
where $\ket{0}$ denotes the non-degenerate ground state and the excited level is $(2^n-1)$-degenerate with an energy $x$ determined by the equation
$e^{x} = \dfrac{(2^n - 1)(x + 2)}{x - 2}$. For $n \to \infty$, it yields the theoretical bound 
\begin{equation}\label{optC_n_large}
    C(\accentset{\ast}{H}_{n,n})\approx\dfrac{n^2(\ln 2)^2}{4},
\end{equation}
with $\beta x=n \ln 2$, which resembles a Heisenberg-like quadratic scaling. For an arbitrary $n$, the optimal Hamiltonian \eqref{eq: H_opt} can be simulated exactly by $n$-local Hamiltonians~\cite{Mok2021}.

Since $n$-body interactions are challenging to engineer experimentally, the optimal thermometric probe was investigated under the more realistic restriction to two-body interactions~\cite{Abiuso_2024,Mok2021,PhysRevA.111.052216}. Particularly, Ref.~\cite{Abiuso_2024} uses numerical optimization to identify  the 2-local \textit{Star model} as an optimal architecture within the  manifold of 2-local Hamiltonians, for $n\geq 6$. The 2-local star model is defined as
\begin{equation}\label{2localStar}
H_{n,2}^{\mathrm{Star}} \coloneqq a Z_1 + b \sum_{i=2}^{n} Z_i (\mathds{1} + Z_1),
\end{equation}
with $a,b \in \mathbb{R}$. In this configuration, a central qubit is coupled to the remaining $n-1$ qubits via pairwise interactions. This model further reaches the mathematical bound $ C(\accentset{\ast}{H}_{n,n})$ 
in the large $n$ limit. For $n<6$, numerical optimization instead found the 2-local \textit{All-to-All model} to be the optimal architecture, featuring uniform pairwise interactions among all qubits:
\begin{equation}\label{2localAtA}
     H_{n,2}^{\mathrm{AtA}}= -h\sum_{i=1}^n Z_i  -J\sum_{i<j}^n Z_i Z_j,
\end{equation}
with $h,J>0$. Numerical optimization up to \mbox{$n=9$} revealed no advantage from including arbitrary 2-local interactions involving also  $X$- and $Y$-Pauli terms, allowing  to identify the Star and All-to-All models as candidates for global optimality over the full class of 2-local Hamiltonians depending on the system size~\cite{Abiuso_2024}.

In this work, we address the problem in full generality. To this aim,  for a family of Hamiltonians ${\bm H}_{n,k}^{\chi}$, we define $\varepsilon ({\bm H}_{n,k}^{\chi})$ as the relative distance of their maximum heat capacity to that of the optimal thermometer, that is: 
\begin{equation}
\varepsilon(\boldsymbol{H}^\chi_{n,k}) \coloneqq \left| \frac{C(\accentset{\ast}{H}_{n,n}) - C(\accentset{\ast}{H}^\chi_{n,k})}{C(\accentset{\ast}{H}_{n,n})} \right|,
\end{equation}
and it holds that $0 \leq \varepsilon(\boldsymbol{H}^\chi_{n,k}) \leq 1$. Thus, the closer the distance $\varepsilon(\boldsymbol{H}^\chi_{n,k}) $ to zero, the better the performance of the class. Throughout, we take $\varepsilon(\boldsymbol{H}^\chi_{n,k})$ as our figure of merit.

\section{Optimal thermometers with  $3$-local Hamiltonians}\label{3local}
\begin{figure}[ht] \centering 
\begin{minipage}{\linewidth} \centering
\includegraphics[width=\linewidth]{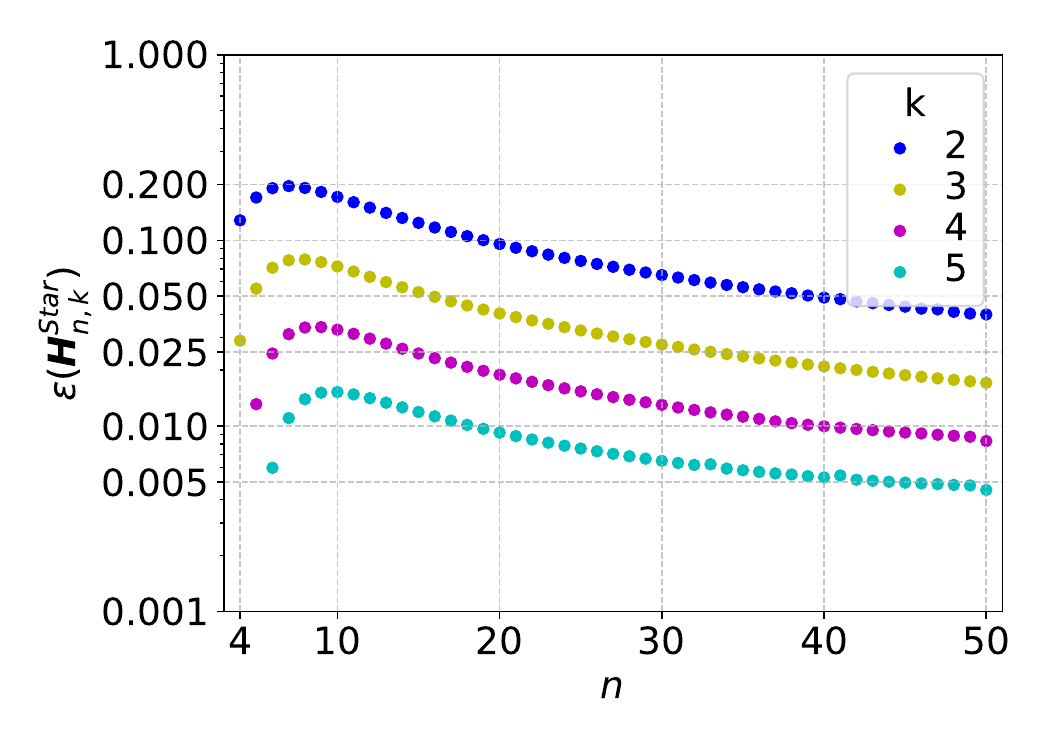}
\caption*{(a)}
\end{minipage} \hfill 

\begin{minipage}{\linewidth} \centering 
\includegraphics[width=\linewidth]{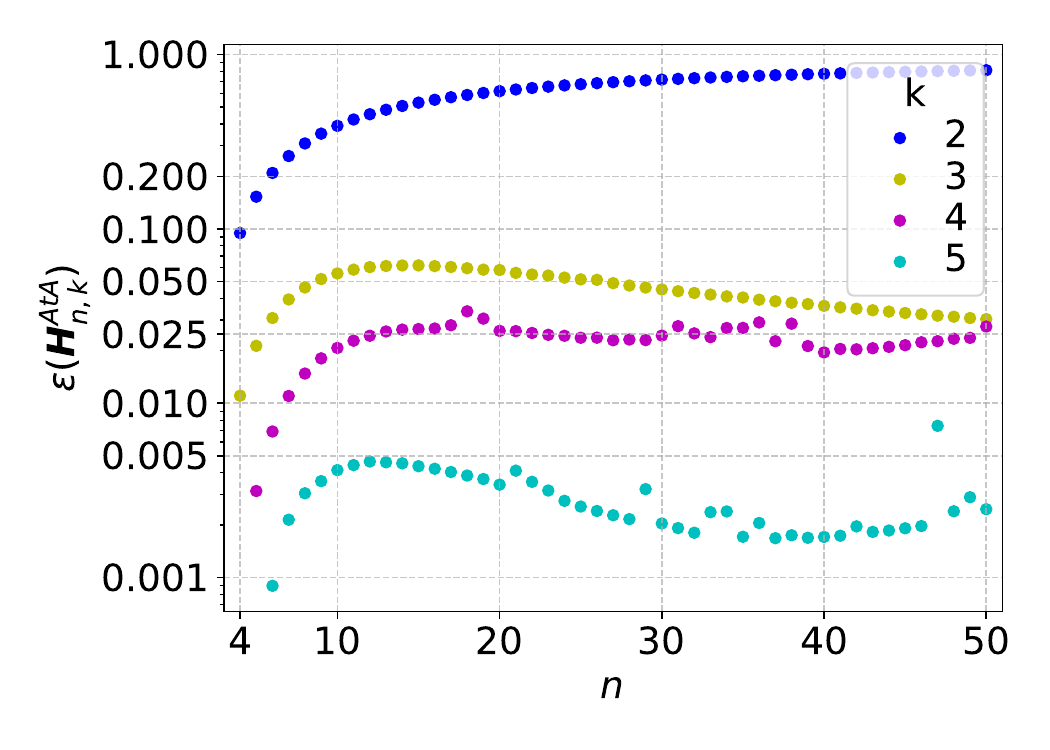}
\caption*{(b)}
\end{minipage}

\captionsetup{justification=Justified, singlelinecheck=false}
\caption{Relative distance of the heat capacity from the theoretical maximum ($\varepsilon({\boldsymbol H}_{n,k}^{\chi})$) as a function of the number of qubits $n$ for $k$-local Hamiltonians with different values of $k$ (Note the logarithmic scale in the vertical axis). The panel (a) corresponds to the Star model (see Eqs.~\eqref{2localStar},~\eqref{eq:3local_star}, and \eqref{eq:H_Star}). Panel (b) corresponds to the All-to-All model (see Eqs.~\eqref{2localAtA},~\eqref{eq:3loc_ata_3_par}, and \eqref{eq:ATA_general}). The observed non-smoothness  results from the optimization converging to local rather than global optima at isolated system sizes.}
\label{fig:relative_error_starAtA} 
\end{figure}

While the 2-local Star model achieves the optimal heat capacity in the limit $n\gg 1$, that is, ${\varepsilon({\bm  H}_{2,n}^{\rm Star}}) \overset{n\gg 1}{\longrightarrow} 0$, as established analytically in~\cite{Abiuso_2024}, finite-size effects remain substantial. For small system sizes, both the 2-local Star and the 2-local All-to-All models exhibit heat capacities well below the global optimum attainable with unrestricted Hamiltonians. 
 In Fig.~\ref{fig:relative_error_starAtA}  (blue dots) we illustrate it for the 2-local All-to-All and the 2-local Star models, along with other $k$-local models. 
While 2-local Hamiltonians are the most straightforward ones to implement in practice, it is natural to ask how and to what extent a modest increase in  $k$-locality can improve thermometry precision. To this aim, we search for $\accentset{\ast}{H}_{n,3}$, and explore its thermometric performance.

The optimization of the heat capacity over the space of $n$-qubit $k$-local Hamiltonians is a highly nonconvex problem. Evaluating the objective function requires diagonalizing the Hamiltonian at each optimization step, while the number of independent parameters grows rapidly with both system size $n$ and locality $k$. In particular, the dimension of the search space scales as $\sum_{\ell=0}^{k} 3^\ell \binom{n}{\ell}$, corresponding to all Pauli strings of weight at most $k$. %it starts in 0 since we also include the term acting with identity on every party
As a consequence, the optimization becomes increasingly computationally demanding for larger systems. For these reasons, throughout this work we employ the Adam optimizer \cite{adam}, a widely used optimization method originally developed for large-scale machine-learning applications. Moreover, owing to the nonconvex nature of the problem, numerical optimization generally provides no guarantee of reaching the global optimum;  nonetheless, it always serves as a lower bound on the achievable heat capacity within a class of Hamiltonians. Additional details of the numerical implementation are provided in Appendix~\ref{adam_details}.

Analogously to the 2-local case, we first optimize over the class of general 3-local Hamiltonians. For system sizes up to $n=8$, the largest instance that could be handled numerically, allowing arbitrary Pauli terms yields no advantage over restricting the optimization to the classical ones. In this regime, the optimal solution is given by the \textit{3-local All-to-All model}:
\begin{equation} \label{eq:3loc_ata_3_par}H_{n,3}^\text{AtA} = a \sum_{i=1}^{n} Z_i + b \sum_{i<j} Z_iZ_j + c \sum_{i<j<k} Z_iZ_jZ_k, 
\end{equation}
with $a,b,c\in\mathbb{R}$.

Motivated by this observation, we conjecture that the optimum for larger systems is also attained within the class of classical Hamiltonians and therefore restrict the optimization accordingly. Within this restriction we solve systems of sizes up to $n=11$.
The 3-local All-to-All model remains optimal throughout this range. For $n>11$, however, the optimization over the full class of classical Hamiltonians also becomes computationally prohibitive. We therefore restrict the search to the $3$-local All-to-All model~(\ref{eq:3loc_ata_3_par}), and a particular 3-local generalization of the  Star model, which we define below. As will be shown later, the spectrum of these models can be found analytically. The objective function can therefore be evaluated without repeatedly evaluating the spectrum during the optimization procedure, making the approach scalable to arbitrary $n$. 

In particular, we consider the following\textit{ 3-local Star model}:
\begin{eqnarray}\label{eq:3local_star}
  H_{n,3}^\text{Star} &=& \alpha_{1} Z_{1} + \alpha_{2} Z_{2} + \alpha_{3} \sum_{i=3}^{n} Z_{i} \nonumber\\
  &+& \alpha_{4} Z_{1} Z_{2}    
  +\alpha_{5} \sum_{i=3}^{n} Z_{1} Z_{i} + \alpha_{6} \sum_{i=3}^{n} Z_{2} Z_{i} \nonumber \\
  &+& \alpha_{7} \sum_{i=3}^{n} Z_{1} Z_{2} Z_{i},  
\end{eqnarray} 
with $\alpha_i\in \mathbb{R}$, $\forall i$. This is inspired by the standard 2-local Star model: since the 2-local Star model describes a distinguished spin coupled uniformly to the remaining $n-1$ spins (see Fig.~\ref{fig:scheme_star}a), a natural 3-local generalization is  obtained by introducing a second distinguished spin. The resulting model includes interactions between the two central spins, as well as two- and three-body couplings involving the remaining $n-2$ spins. Allowing the two central spins to couple differently to the rest of the system makes $H_{n,3}^\text{Star}$ the most general Hamiltonian of this class (see Fig.~\ref{fig:scheme_star}b). Note that the 2-local Star model is recovered as a particular case of the 3-local Star Hamiltonian by setting $\alpha_1 = a$, $\alpha_2 = \alpha_3 = \alpha_4 = \alpha_5 = b$, $\alpha_6 = \alpha_7 = 0$.
\begin{figure}[t] \centering \begin{minipage}{0.48\textwidth} \centering \begin{tikzpicture} 
% Nodo central (mismo tamaño que los nodos b)
\node[circle, draw=orange!80!black, fill=orange!70, inner sep=3pt] (a) at (0,0) {};
\node at (0,0.4) {$a$}; % etiqueta afuera

% Nodos periféricos y conexiones
\foreach \i in {0,1,...,8} {
    \node[circle, draw=blue!80, fill=black, inner sep=3pt] (b\i) at (60*\i:2cm) {};
    \draw[thick, gray] (a) -- node[midway, above left, sloped] {$b$} (b\i);
    \node at ($(b\i) + (60*\i:0.6cm)$) {$b$};
}

\end{tikzpicture} \caption*{(a)} \end{minipage} \hfill \begin{minipage}{0.48\textwidth} \centering \begin{tikzpicture} 
\tikzset{
nodeA/.style={circle, draw=orange, fill=orange!70, inner sep=3pt},
nodeB/.style={circle, draw=black, fill=black!80, inner sep=3pt},
edge/.style={thick, gray},
triple/.style={thick, red, bend left=20},
}

% Nodos centrales
\node[nodeA] (a1) at (-1,0) {};
\node[below=1pt of a1] {$\alpha_1$};

\node[nodeA] (a2) at (1,0) {};
\node[below=1pt of a2] {$\alpha_2$};

\foreach \i in {0,1,...,8} {
\node[circle, draw=blue!80, fill=black, inner sep=3pt] (b\i) at (60*\i:2cm) {};
\node at ($(b\i) + (60*\i:0.6cm)$) {$\alpha_3$};
}

% Conexiones a 2 cuerpos (gris)
\draw[edge] (a1) -- (a2) node[midway, above] {$\alpha_4$};
\draw[edge] (a1) -- (b1) node[sloped, midway, below=0.000001cm] {$\alpha_5$};
\draw[edge] (a2) -- (b1) node[sloped, midway, above] {$\alpha_6$};

% Conexiones a 3 cuerpos (rojo)
\draw[triple] (a1) to[bend left=25] node[pos=0.4, left] {$\alpha_7$} (b1);
\draw[triple] (b1) to[bend left=25] node[pos=0.6, right] {$\alpha_7$} (a2);
\draw[triple] (a2) to[bend left=25] node[pos=0.5, below] {$\alpha_7$} (a1);

\end{tikzpicture} \caption*{(b)} \end{minipage} \captionsetup{justification=Justified, singlelinecheck=false}\caption{Schematic representation of the Star models: (a) 2-local and (b) 3-local. Gray lines represent two-body interactions, while red lines represent three-body interactions. For visual clarity, in the 3-local case, only the interactions with one of the non-privileged spins are shown.}
\label{fig:scheme_star}
\end{figure}
An alternative 3-local generalization of the Star model was also investigated; however, it yielded a lower maximal heat capacity (see Appendix~\ref{alternative_star}). Moreover, upon optimizing the parameters \(\alpha_i\), we find that the optimal coefficients satisfy the relations 
\begin{equation}\label{eq:opt_3loc}
\alpha_1=\alpha_2=\alpha_4, \qquad \alpha_3=\alpha_5=\alpha_6=\alpha_7. \end{equation}

Fig.~\ref{fig:relative_error_starAtA}  shows the relative errors achieved by the 3-local All-to-All and 3-local Star models (yellow dots).
We first observe that the 3-local All-to-All and Star models significantly outperform their 2-local counterparts, showing that even a minimal increase in locality can yield a substantial improvement in thermometric performance. Furthermore, we find that the heat capacity of the \mbox{3-local} Star model converges to the theoretical optimum in the asymptotic limit, a result that is indeed expected: since  \mbox{$C(\accentset{\ast}{H}_{n,n})\geq C(  \accentset{\ast}{H}_{n,2}^{\text{Star}})\geq C(  \accentset{\ast}{H}_{n-1,n-1})$}~\cite{Abiuso_2024}, and given that ${\boldsymbol H}_{n,2}^{\rm Star} \subset {\boldsymbol H}_{n,3}^{\rm Star}$,  it follows that  \mbox{$C(  \accentset{\ast}{H}_{n,n})\geq C(  \accentset{\ast}{H}_{n,3}^\text{Star}) \geq C(  \accentset{\ast}{H}_{n,2}^{\text{Star}})\geq C(  \accentset{\ast}{H}_{n-1,n-1})$}. In the limit of $n \gg 1$, by dividing everything by $C(  \accentset{\ast}{H}_{n,n})$, one gets ${\varepsilon}({\boldsymbol H}_{n,3}^{\rm Star}) \approx {\varepsilon}({\boldsymbol H}_{n,2}^{\rm Star}) \approx 0$.

Lastly, we compare the two proposed 3-local models here in Fig.~\ref{fig: error_Star_vs_AtA_v2}.
The figure displays the difference between the relative errors of the Star and the All-to-All models (yellow dots). A positive (negative) value indicates that the All-to-All (Star) model performs better than the Star (All-to-All) model. 
We observe that the  All-to-All model performs better up to $n=12$, while the Star model exhibits a larger heat capacity for $n>12$.

\section{Optimal thermometers with $k$-local All-to-All and Star Hamiltonians}\label{klocal}

We wonder if the enhanced performance observed by increasing $k$-locality from $2$ to $3$ persists at higher $k$. In this section, we generalize the models introduced above to arbitrary $k$-local interactions and derive their energy spectra to elucidate how the resulting spectral structure controls the heat capacity. We then analyze the optimal heat capacity achievable as the locality $k$ is increased.

\subsection{$k$-local All-to-All model} \label{sec:k_local_AtA}

The All-to-All model~\eqref{eq:3loc_ata_3_par}
admits a natural generalization to arbitrary $k$-local interactions:

\begin{equation}\label{eq:ATA_general}
    H_{n,k}^{\mathrm{AtA}}= \sum_{q=1}^k\alpha _q \sum_{\substack{S \subseteq \{1,\dots,n\} \\   \left| S \right|=q}}
\prod_{i \in S} Z_i,
\end{equation}
where $\alpha_q\in\mathbb{R}$ and $S$ runs over all subsets of $q$ qubits.

The energy levels of the $k$-local All-to-All Hamiltonian can be obtained analytically. Identifying the computational-basis states of each qubit with spin-up and spin-down states according to $Z_i=\pm 1$, we label each configuration by the number $p$ of spins up. For a configuration with $p$ spins up, the corresponding eigenvalue reads
\begin{equation}
E_p =
\sum_{q=1}^{k} \alpha_q 
\sum_{j=0}^{q}
\binom{p}{j}
\binom{n-p}{q-j}
(-1)^{\,q-j},
\end{equation}
each level with degeneracy $\binom{n}{p}$. 
In Appendix \ref{app:energy_derivation} it is shown that the model converges to the theoretical optimum in the fully $n$-local limit.
Indeed, for $k=n$ and constant coefficients $\alpha_q = a < 0$, one obtains 

\begin{equation}
E_p =
\begin{cases}
(2^n - 1)a, & p = n, \\
-a, & 0 \le p < n,
\end{cases}
\end{equation}
thus recovering the globally optimal spectrum consisting of a single non-degenerate level separated from a \mbox{$(2^n-1)$-degenerate} level. As shown in Fig.~\ref{fig:relative_error_starAtA}a, the relative deviation decreases monotonically with increasing locality $k$, indicating that higher-locality interactions bring the heat capacity progressively closer to the theoretical maximum, which is ultimately attained at $k=n$. Expectedly, the improvement becomes progressively less pronounced as $k$ increases.

\begin{figure}[t]
 \centering
  \includegraphics[width=0.9\linewidth]{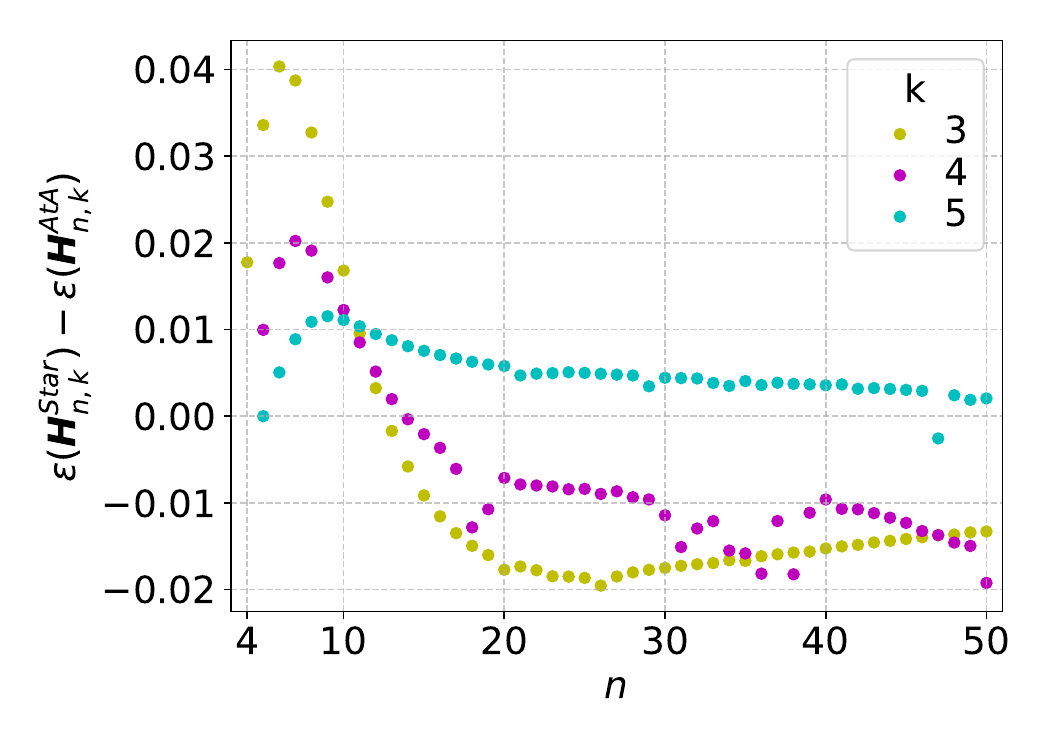}

  \captionsetup{justification=Justified, singlelinecheck=false}\caption{
Difference between the relative errors of the $k$-local Star and $k$-local All-to-All models as a function of the number of qubits $n$, for different values of $k$.
}
  \label{fig: error_Star_vs_AtA_v2}
\end{figure}

\subsection{$k$-local Star model}

The $k$-local Star model, comprising $k-1$ central spins and $n-k+1$ satellite spins, is defined as
\begin{equation}
H_{n,k}^{\mathrm{Star}} =
a \sum_{\substack{S \subseteq \{1,\dots,k-1\} \\ S \neq \emptyset}}
\prod_{i \in S} Z_i
+ b \sum_{r=k}^{n} Z_r \prod_{i=1}^{k-1} (\mathds{1} + Z_i),
\label{eq:H_Star}
\end{equation}
where $a,b\in \mathbb{R}$ and $S$ runs over all nonempty subsets of the $k-1$ central spins. Note that the optimal 3-local Star Hamiltonian, despite its more general parametrization, is characterized by only two independent coefficients~\eqref{eq:opt_3loc}. We therefore extend this two-parameter structure to arbitrary locality $k$. It is also worth noting that the first term of the Star formula is equivalent to an all-to-all Hamiltonian of $k-1$ spins with $(k-1)$-body interactions, meaning that the $k$-local Star model features an all-to-all interaction among the $k-1$ central spins.

\begin{figure}[ht]
\centering
\begin{subfigure}{0.5\columnwidth}
    \centering
    \includegraphics[width=\linewidth]{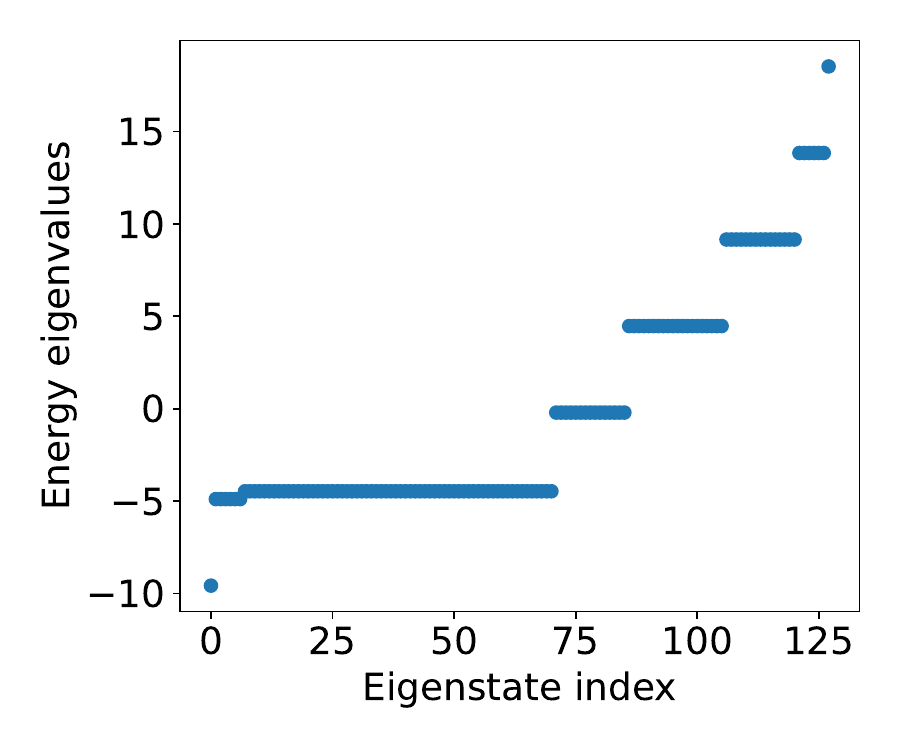}
    \caption{$k=2$}
\end{subfigure}\hfill
\begin{subfigure}{0.5\columnwidth}
    \centering
    \includegraphics[width=\linewidth]{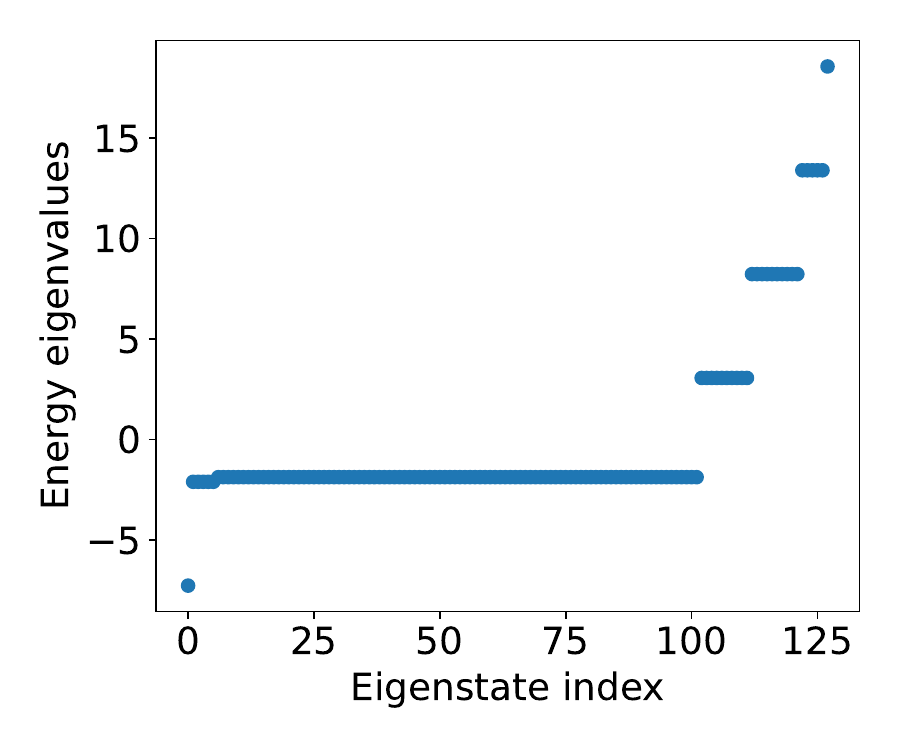}
    \caption{$k=3$}
\end{subfigure}

\medskip

\begin{subfigure}{0.5\columnwidth}
    \centering
    \includegraphics[width=\linewidth]{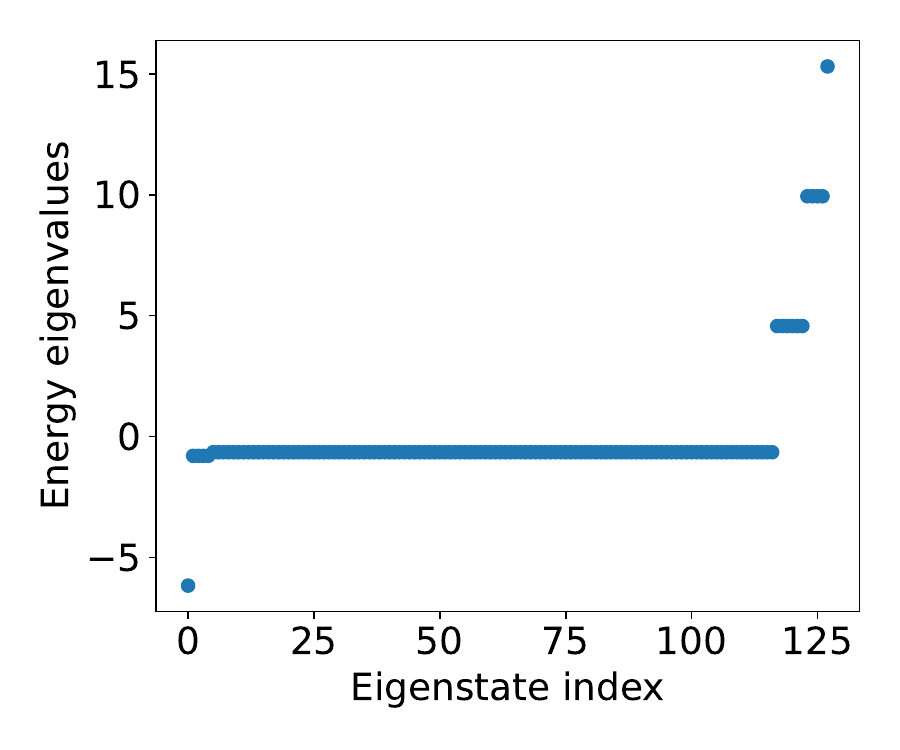}
    \caption{$k=4$}
\end{subfigure}\hfill
\begin{subfigure}{0.5\columnwidth}
    \centering
    \includegraphics[width=\linewidth]{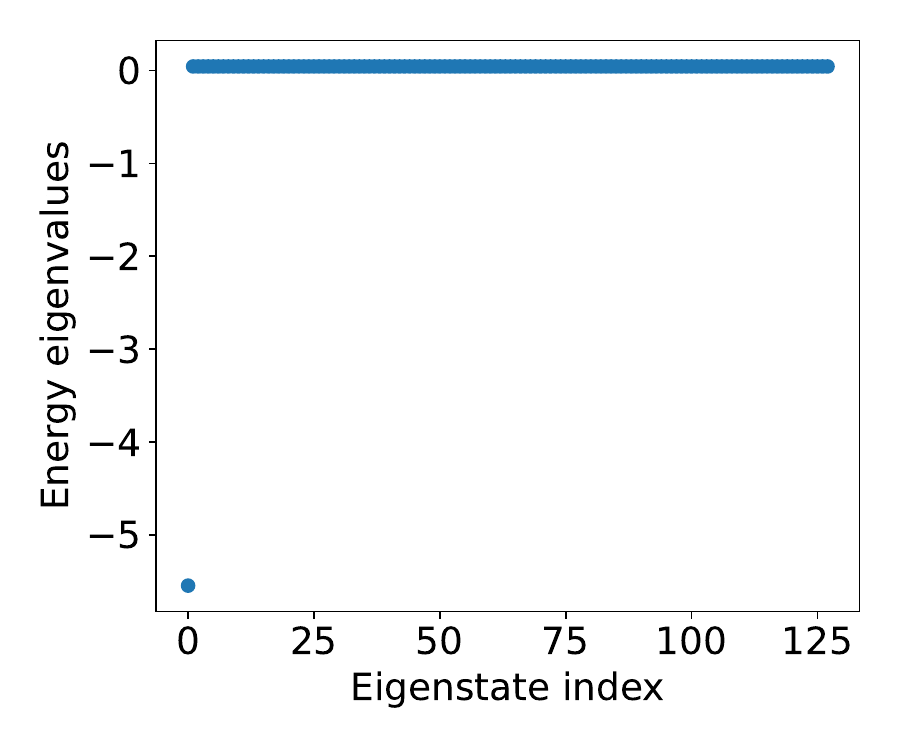}
    \caption{$k=7$}
\end{subfigure}

\captionsetup{justification=Justified, singlelinecheck=false}\caption{Eigenvalue spectra of the optimized 7-qubit $k$-local Star Hamiltonian, for different values of $k$.}
\label{fig:spectra7qubits}
\end{figure}

The spectrum of this Hamiltonian can be also obtained analytically. It consists of a highly degenerate level
\begin{equation}
E_{\mathrm{deg}} = -a,
\label{eq:Edeg_star}
\end{equation}
with degeneracy $2^n - 2^{\,n-k+1}$, and a family of excited levels
\begin{equation}
E_p = (2^{k-1}-1)a + 2^{k-1}b(2p + (k-1) - n), 
\label{eq:Ep_star}
\end{equation} where each level corresponds to $p$ satellite spins up with a degeneracy of $\binom{n-k+1}{p}$. The spectral structure reveals how locality controls thermometric performance, as is clearly illustrated in Fig.~\ref{fig:spectra7qubits}, where we show the optimized spectra for $n=7$ and different values of $k$. 
As $k$ increases, the degeneracy of the highly degenerate level grows, while the degeneracy of the higher excited levels decreases.  For $k=n$ and choosing $a=-b$, it is easy to see that the spectrum is reduced to $E_0 = (2^n-1)a$,
$E_{\mathrm{deg}} = -a$, 
ultimately reproducing the globally optimal spectrum, characterized by a single nondegenerate level and a highly degenerate level. This is in analogy to the All-to-All model. Indeed, numerical optimization shows that the thermometric performance improves with increasing locality $k$, albeit with diminishing gains as $k$ increases, reaching the theoretical maximum in the fully $n$-local case (see Fig.~\ref{fig:relative_error_starAtA}b). A similar argument to the one we made for the 3-local scenario guarantees that \mbox{$C( \accentset{\ast}{H}_{n,n})\geq C( \accentset{\ast}{H}_{n,k}^{\text{Star}}) \geq C( \accentset{\ast}{H}_{n,2}^{\text{Star}})\geq C( \accentset{\ast}{H}_{n-1,n-1})$} and thus for $n\gg 1$,  ${\varepsilon}(\boldsymbol{H}_{n,k}^{\rm Star}) \approx 0$.

Finally, in Fig.~\ref{fig: error_Star_vs_AtA_v2} we plot the difference between the relative errors of the
$k$-local Star and $k$-local All-to-All models as a function
of the number of qubits $n$, for \mbox{$k=4,5$}. Isolated deviations from the overall trend are likely numerical artifacts of the optimization procedure, which may have converged to local rather than global optima for these system sizes.

\section{Discussion}\label{discussion}

In this work, we have investigated the thermometric performance of equilibrium $n$-qubit quantum probes as a function of Hamiltonian locality and identified configurations that substantially enhance the heat capacity at finite system sizes. We introduced new families of $3$-local Hamiltonians, namely the All-to-All model \eqref{eq:ATA_general} and Star model \eqref{eq:H_Star}. 
To begin with, for $n\leq 8$, we numerically prove optimality of 3-local All-to-All Hamiltonians among any 3-local Hamiltonian. They further show a significantly better performance than the optimal 2-local ones.
Inspired by the optimality of these classical models, for $n>8$ we restrict our numerical analysis to classical Hamiltonians to bypass computational limitations.
Within this set we observe that among 3-local Hamiltonians the 3-local All-to-All model is optimal when $n\leq 11$. For $n>11$, the optimization was further restricted to the 3-local All-to-All-Model and the 3-local Star models, with the latter consistently yielding a higher heat capacity except for $n=12$.  Notably, the 3-local Star model---which contains the 2-local Star model as a special case---retains the optimal quadratic scaling in the thermodynamic limit~\cite{Abiuso_2024}, showing that finite-size performance can be improved without compromising the asymptotic scaling. We then put forward the $k$-local extensions of the All-to-All and Star models, and showed that increasing locality progressively brings their thermometric performance closer to the unrestricted theoretical optimum. Analysis of the corresponding energy spectra reveals that this enhancement is accompanied by a gradual convergence toward the spectral structure of the optimal thermometer. Expectedly, the improvement becomes progressively less pronounced as the optimal structure is approached. 

Our analysis opens the door to use the Fisher information (heat capacity) as a witness of $k$-locality of a quantum system with an unknown Hamiltonian $H$. 
Namely, if ${\varepsilon}(H) < {\varepsilon}(\accentset{\ast}{\boldsymbol H}_{n,k})$, then $H$ is at least $(k+1)$-local.
While finding $\accentset{\ast}{\boldsymbol H}_{n,k}$ remains a big challenge for general $n$ and $k$, our results partly answer this question. Namely, using  Fig.~\ref{fig:relative_error_starAtA} as a guide, for $n\leq 8 $ any system with ${\varepsilon}(H) < {\varepsilon}(\accentset{\ast}{\boldsymbol H}_{n,3}^{\rm AtA}) $ is at least $4$-local. Furthermore, due to the results of Ref~\cite{Abiuso_2024}, for $n < 6$ if ${\varepsilon}(H) < {\varepsilon}(\accentset{\ast}{\boldsymbol H}_{n,2}^{\rm AtA}) $ then $H$ is at least $3$-local. 

Our results therefore establish a direct connection between Hamiltonian locality, spectral structure, and thermometric performance. They show that few-body interactions can capture much of the advantage of unrestricted Hamiltonians, thereby offering a promising route for the design of quantum thermometers.

%An important direction for future work is to rigorously establish  the optimality of the All-to-All model at small system sizes and to derive the optimal structure for larger systems. Such analytical results are particularly desirable since numerical methods cannot guarantee global optimality and the optimization becomes computationally inaccessible as the system size increases. Beyond overcoming these limitations, they could provide deeper insight into the relation between locality and thermometric performance.

\section*{Acknowledgements}
Discussions with Martí Perarnau-Llobet and Paolo Andrea Erdman are gratefully appreciated. J.L.G. was supported by  Collaboration Scholarships from the Universidad Politécnica de Madrid and the Ministerio de Educación, Formación Profesional y Deportes. M.G.D. acknowledges financial support from the following grants: PID2020-113523GB-I00 and PID2023-146758NB-I00 funded by MICIU/AEI/10.13039/501100011033. A.S. acknowledges PID2022-141283NB-I00 and the SGR grant from Generalitat de Catalunya for financial support. The authors acknowledge TU Wien Bibliothek for financial support through its Open Access Funding Programme. 
This research was funded in part by the Austrian Science Fund (FWF) [“NOQUS,” Grant DOI: 10.55776/PAT1969224].
%
%Microsoft Copilot was used in editing text.

{\bf Data availability---}All the codes are available on this github repository: \url{https://github.com/jorge-lopez-gonzalez/Optimal_Quantum_Thermometers_with_k-local_Hamiltonians}

\bibliography{bibliografia}

\begin{thebibliography}{42}%
\makeatletter
\providecommand \@ifxundefined [1]{%
 \@ifx{#1\undefined}
}%
\providecommand \@ifnum [1]{%
 \ifnum #1\expandafter \@firstoftwo
 \else \expandafter \@secondoftwo
 \fi
}%
\providecommand \@ifx [1]{%
 \ifx #1\expandafter \@firstoftwo
 \else \expandafter \@secondoftwo
 \fi
}%
\providecommand \natexlab [1]{#1}%
\providecommand \enquote  [1]{``#1''}%
\providecommand \bibnamefont  [1]{#1}%
\providecommand \bibfnamefont [1]{#1}%
\providecommand \citenamefont [1]{#1}%
\providecommand \href@noop [0]{\@secondoftwo}%
\providecommand \href [0]{\begingroup \@sanitize@url \@href}%
\providecommand \@href[1]{\@@startlink{#1}\@@href}%
\providecommand \@@href[1]{\endgroup#1\@@endlink}%
\providecommand \@sanitize@url [0]{\catcode `\\12\catcode `\$12\catcode
  `\&12\catcode `\#12\catcode `\^12\catcode `\_12\catcode `\%12\relax}%
\providecommand \@@startlink[1]{}%
\providecommand \@@endlink[0]{}%
\providecommand \url  [0]{\begingroup\@sanitize@url \@url }%
\providecommand \@url [1]{\endgroup\@href {#1}{\urlprefix }}%
\providecommand \urlprefix  [0]{URL }%
\providecommand \Eprint [0]{\href }%
\providecommand \doibase [0]{https://doi.org/}%
\providecommand \selectlanguage [0]{\@gobble}%
\providecommand \bibinfo  [0]{\@secondoftwo}%
\providecommand \bibfield  [0]{\@secondoftwo}%
\providecommand \translation [1]{[#1]}%
\providecommand \BibitemOpen [0]{}%
\providecommand \bibitemStop [0]{}%
\providecommand \bibitemNoStop [0]{.\EOS\space}%
\providecommand \EOS [0]{\spacefactor3000\relax}%
\providecommand \BibitemShut  [1]{\csname bibitem#1\endcsname}%
\let\auto@bib@innerbib\@empty
%</preamble>
\bibitem [{\citenamefont {Giazotto}\ \emph {et~al.}(2006)\citenamefont
  {Giazotto}, \citenamefont {Heikkil{\"a}}, \citenamefont {Luukanen},
  \citenamefont {Savin},\ and\ \citenamefont {Pekola}}]{Giazotto2006}%
  \BibitemOpen
  \bibfield  {author} {\bibinfo {author} {\bibfnamefont {F.}~\bibnamefont
  {Giazotto}}, \bibinfo {author} {\bibfnamefont {T.~T.}\ \bibnamefont
  {Heikkil{\"a}}}, \bibinfo {author} {\bibfnamefont {A.}~\bibnamefont
  {Luukanen}}, \bibinfo {author} {\bibfnamefont {A.~M.}\ \bibnamefont
  {Savin}},\ and\ \bibinfo {author} {\bibfnamefont {J.~P.}\ \bibnamefont
  {Pekola}},\ }\bibfield  {title} {\bibinfo {title} {Opportunities for
  mesoscopics in thermometry and refrigeration: Physics and applications},\
  }\href {https://doi.org/10.1103/RevModPhys.78.217} {\bibfield  {journal}
  {\bibinfo  {journal} {Reviews of Modern Physics}\ }\textbf {\bibinfo {volume}
  {78}},\ \bibinfo {pages} {217} (\bibinfo {year} {2006})}\BibitemShut
  {NoStop}%
\bibitem [{\citenamefont {Yue}\ and\ \citenamefont {Wang}(2012)}]{Yue2012}%
  \BibitemOpen
  \bibfield  {author} {\bibinfo {author} {\bibfnamefont {Y.}~\bibnamefont
  {Yue}}\ and\ \bibinfo {author} {\bibfnamefont {X.}~\bibnamefont {Wang}},\
  }\bibfield  {title} {\bibinfo {title} {Nanoscale thermal probing},\ }\href
  {https://doi.org/10.3402/nano.v3i0.11586} {\bibfield  {journal} {\bibinfo
  {journal} {Nano Reviews}\ }\textbf {\bibinfo {volume} {3}},\ \bibinfo {pages}
  {11586} (\bibinfo {year} {2012})}\BibitemShut {NoStop}%
\bibitem [{\citenamefont {Mehboudi}\ \emph
  {et~al.}(2019{\natexlab{a}})\citenamefont {Mehboudi}, \citenamefont
  {Sanpera},\ and\ \citenamefont {Correa}}]{Mehboudi2019}%
  \BibitemOpen
  \bibfield  {author} {\bibinfo {author} {\bibfnamefont {M.}~\bibnamefont
  {Mehboudi}}, \bibinfo {author} {\bibfnamefont {A.}~\bibnamefont {Sanpera}},\
  and\ \bibinfo {author} {\bibfnamefont {L.~A.}\ \bibnamefont {Correa}},\
  }\bibfield  {title} {\bibinfo {title} {Thermometry in the quantum regime:
  Recent theoretical progress},\ }\href
  {https://doi.org/10.1088/1751-8121/ab2828} {\bibfield  {journal} {\bibinfo
  {journal} {Journal of Physics A: Mathematical and Theoretical}\ }\textbf
  {\bibinfo {volume} {52}},\ \bibinfo {pages} {303001} (\bibinfo {year}
  {2019}{\natexlab{a}})}\BibitemShut {NoStop}%
\bibitem [{\citenamefont {De~Pasquale}\ and\ \citenamefont
  {Stace}(2018)}]{DePasquale2018}%
  \BibitemOpen
  \bibfield  {author} {\bibinfo {author} {\bibfnamefont {A.}~\bibnamefont
  {De~Pasquale}}\ and\ \bibinfo {author} {\bibfnamefont {T.~M.}\ \bibnamefont
  {Stace}},\ }\bibfield  {title} {\bibinfo {title} {Quantum thermometry},\ }in\
  \href {https://doi.org/10.1007/978-3-319-99046-0_21} {\emph {\bibinfo
  {booktitle} {Thermodynamics in the Quantum Regime}}},\ \bibinfo {series}
  {Fundamental Theories of Physics}, Vol.\ \bibinfo {volume} {195}\ (\bibinfo
  {publisher} {Springer International Publishing},\ \bibinfo {address} {Cham},\
  \bibinfo {year} {2018})\ pp.\ \bibinfo {pages} {503--527}\BibitemShut
  {NoStop}%
\bibitem [{\citenamefont {Kucsko}\ \emph {et~al.}(2013)\citenamefont {Kucsko},
  \citenamefont {Maurer}, \citenamefont {Yao}, \citenamefont {Kubo},
  \citenamefont {Noh}, \citenamefont {Lo}, \citenamefont {Park},\ and\
  \citenamefont {Lukin}}]{Kucsko2013}%
  \BibitemOpen
  \bibfield  {author} {\bibinfo {author} {\bibfnamefont {G.}~\bibnamefont
  {Kucsko}}, \bibinfo {author} {\bibfnamefont {P.~C.}\ \bibnamefont {Maurer}},
  \bibinfo {author} {\bibfnamefont {N.~Y.}\ \bibnamefont {Yao}}, \bibinfo
  {author} {\bibfnamefont {M.}~\bibnamefont {Kubo}}, \bibinfo {author}
  {\bibfnamefont {H.~J.}\ \bibnamefont {Noh}}, \bibinfo {author} {\bibfnamefont
  {P.~K.}\ \bibnamefont {Lo}}, \bibinfo {author} {\bibfnamefont
  {H.}~\bibnamefont {Park}},\ and\ \bibinfo {author} {\bibfnamefont {M.~D.}\
  \bibnamefont {Lukin}},\ }\bibfield  {title} {\bibinfo {title}
  {Nanometre-scale thermometry in a living cell},\ }\href
  {https://doi.org/10.1038/nature12373} {\bibfield  {journal} {\bibinfo
  {journal} {Nature}\ }\textbf {\bibinfo {volume} {500}},\ \bibinfo {pages}
  {54} (\bibinfo {year} {2013})}\BibitemShut {NoStop}%
\bibitem [{\citenamefont {Fujiwara}\ \emph {et~al.}(2020)\citenamefont
  {Fujiwara}, \citenamefont {Sun}, \citenamefont {Dohms}, \citenamefont
  {Nishimura}, \citenamefont {Suto}, \citenamefont {Takezawa}, \citenamefont
  {Oshimi}, \citenamefont {Zhao}, \citenamefont {Sadzak}, \citenamefont
  {Umehara}, \citenamefont {Teki}, \citenamefont {Komatsu}, \citenamefont
  {Benson}, \citenamefont {Shikano},\ and\ \citenamefont
  {Kage-Nakadai}}]{Fujiwara2020}%
  \BibitemOpen
  \bibfield  {author} {\bibinfo {author} {\bibfnamefont {M.}~\bibnamefont
  {Fujiwara}}, \bibinfo {author} {\bibfnamefont {S.}~\bibnamefont {Sun}},
  \bibinfo {author} {\bibfnamefont {A.}~\bibnamefont {Dohms}}, \bibinfo
  {author} {\bibfnamefont {Y.}~\bibnamefont {Nishimura}}, \bibinfo {author}
  {\bibfnamefont {K.}~\bibnamefont {Suto}}, \bibinfo {author} {\bibfnamefont
  {Y.}~\bibnamefont {Takezawa}}, \bibinfo {author} {\bibfnamefont
  {K.}~\bibnamefont {Oshimi}}, \bibinfo {author} {\bibfnamefont
  {L.}~\bibnamefont {Zhao}}, \bibinfo {author} {\bibfnamefont {N.}~\bibnamefont
  {Sadzak}}, \bibinfo {author} {\bibfnamefont {Y.}~\bibnamefont {Umehara}},
  \bibinfo {author} {\bibfnamefont {Y.}~\bibnamefont {Teki}}, \bibinfo {author}
  {\bibfnamefont {N.}~\bibnamefont {Komatsu}}, \bibinfo {author} {\bibfnamefont
  {O.}~\bibnamefont {Benson}}, \bibinfo {author} {\bibfnamefont
  {Y.}~\bibnamefont {Shikano}},\ and\ \bibinfo {author} {\bibfnamefont
  {E.}~\bibnamefont {Kage-Nakadai}},\ }\bibfield  {title} {\bibinfo {title}
  {Real-time nanodiamond thermometry probing in vivo thermogenic responses},\
  }\href {https://doi.org/10.1126/sciadv.aba9636} {\bibfield  {journal}
  {\bibinfo  {journal} {Science Advances}\ }\textbf {\bibinfo {volume} {6}},\
  \bibinfo {pages} {eaba9636} (\bibinfo {year} {2020})}\BibitemShut {NoStop}%
\bibitem [{\citenamefont {Gasparinetti}\ \emph {et~al.}(2015)\citenamefont
  {Gasparinetti}, \citenamefont {Viisanen}, \citenamefont {Saira},
  \citenamefont {Faivre}, \citenamefont {Arzeo}, \citenamefont {Meschke},\ and\
  \citenamefont {Pekola}}]{Gasparinetti2015}%
  \BibitemOpen
  \bibfield  {author} {\bibinfo {author} {\bibfnamefont {S.}~\bibnamefont
  {Gasparinetti}}, \bibinfo {author} {\bibfnamefont {K.~L.}\ \bibnamefont
  {Viisanen}}, \bibinfo {author} {\bibfnamefont {O.-P.}\ \bibnamefont {Saira}},
  \bibinfo {author} {\bibfnamefont {T.}~\bibnamefont {Faivre}}, \bibinfo
  {author} {\bibfnamefont {M.}~\bibnamefont {Arzeo}}, \bibinfo {author}
  {\bibfnamefont {M.}~\bibnamefont {Meschke}},\ and\ \bibinfo {author}
  {\bibfnamefont {J.~P.}\ \bibnamefont {Pekola}},\ }\bibfield  {title}
  {\bibinfo {title} {Fast electron thermometry for ultrasensitive calorimetric
  detection},\ }\href {https://doi.org/10.1103/PhysRevApplied.3.014007}
  {\bibfield  {journal} {\bibinfo  {journal} {Physical Review Applied}\
  }\textbf {\bibinfo {volume} {3}},\ \bibinfo {pages} {014007} (\bibinfo {year}
  {2015})}\BibitemShut {NoStop}%
\bibitem [{\citenamefont {Halbertal}\ \emph {et~al.}(2016)\citenamefont
  {Halbertal}, \citenamefont {Cuppens}, \citenamefont {Shalom}, \citenamefont
  {Embon}, \citenamefont {Shadmi}, \citenamefont {Anahory}, \citenamefont
  {Naren}, \citenamefont {Sarkar}, \citenamefont {Uri}, \citenamefont {Ronen},
  \citenamefont {Myasoedov}, \citenamefont {Levitov}, \citenamefont
  {Joselevich}, \citenamefont {Geim},\ and\ \citenamefont
  {Zeldov}}]{Halbertal2016}%
  \BibitemOpen
  \bibfield  {author} {\bibinfo {author} {\bibfnamefont {D.}~\bibnamefont
  {Halbertal}}, \bibinfo {author} {\bibfnamefont {J.}~\bibnamefont {Cuppens}},
  \bibinfo {author} {\bibfnamefont {M.~B.}\ \bibnamefont {Shalom}}, \bibinfo
  {author} {\bibfnamefont {L.}~\bibnamefont {Embon}}, \bibinfo {author}
  {\bibfnamefont {N.}~\bibnamefont {Shadmi}}, \bibinfo {author} {\bibfnamefont
  {Y.}~\bibnamefont {Anahory}}, \bibinfo {author} {\bibfnamefont {H.~R.}\
  \bibnamefont {Naren}}, \bibinfo {author} {\bibfnamefont {J.}~\bibnamefont
  {Sarkar}}, \bibinfo {author} {\bibfnamefont {A.}~\bibnamefont {Uri}},
  \bibinfo {author} {\bibfnamefont {Y.}~\bibnamefont {Ronen}}, \bibinfo
  {author} {\bibfnamefont {Y.}~\bibnamefont {Myasoedov}}, \bibinfo {author}
  {\bibfnamefont {L.~S.}\ \bibnamefont {Levitov}}, \bibinfo {author}
  {\bibfnamefont {E.}~\bibnamefont {Joselevich}}, \bibinfo {author}
  {\bibfnamefont {A.~K.}\ \bibnamefont {Geim}},\ and\ \bibinfo {author}
  {\bibfnamefont {E.}~\bibnamefont {Zeldov}},\ }\bibfield  {title} {\bibinfo
  {title} {Nanoscale thermal imaging of dissipation in quantum systems},\
  }\href {https://doi.org/10.1038/nature19843} {\bibfield  {journal} {\bibinfo
  {journal} {Nature}\ }\textbf {\bibinfo {volume} {539}},\ \bibinfo {pages}
  {407} (\bibinfo {year} {2016})}\BibitemShut {NoStop}%
\bibitem [{\citenamefont {Karimi}\ \emph {et~al.}(2020)\citenamefont {Karimi},
  \citenamefont {Brange}, \citenamefont {Samuelsson},\ and\ \citenamefont
  {Pekola}}]{Karimi2020}%
  \BibitemOpen
  \bibfield  {author} {\bibinfo {author} {\bibfnamefont {B.}~\bibnamefont
  {Karimi}}, \bibinfo {author} {\bibfnamefont {F.}~\bibnamefont {Brange}},
  \bibinfo {author} {\bibfnamefont {P.}~\bibnamefont {Samuelsson}},\ and\
  \bibinfo {author} {\bibfnamefont {J.~P.}\ \bibnamefont {Pekola}},\ }\bibfield
   {title} {\bibinfo {title} {Reaching the ultimate energy resolution of a
  quantum detector},\ }\href {https://doi.org/10.1038/s41467-019-14247-2}
  {\bibfield  {journal} {\bibinfo  {journal} {Nature Communications}\ }\textbf
  {\bibinfo {volume} {11}},\ \bibinfo {pages} {367} (\bibinfo {year}
  {2020})}\BibitemShut {NoStop}%
\bibitem [{\citenamefont {Hohmann}\ \emph {et~al.}(2016)\citenamefont
  {Hohmann}, \citenamefont {Kindermann}, \citenamefont {Lausch}, \citenamefont
  {Mayer}, \citenamefont {Schmidt},\ and\ \citenamefont
  {Widera}}]{Hohmann2016}%
  \BibitemOpen
  \bibfield  {author} {\bibinfo {author} {\bibfnamefont {M.}~\bibnamefont
  {Hohmann}}, \bibinfo {author} {\bibfnamefont {F.}~\bibnamefont {Kindermann}},
  \bibinfo {author} {\bibfnamefont {T.}~\bibnamefont {Lausch}}, \bibinfo
  {author} {\bibfnamefont {D.}~\bibnamefont {Mayer}}, \bibinfo {author}
  {\bibfnamefont {F.}~\bibnamefont {Schmidt}},\ and\ \bibinfo {author}
  {\bibfnamefont {A.}~\bibnamefont {Widera}},\ }\bibfield  {title} {\bibinfo
  {title} {Single-atom thermometer for ultracold gases},\ }\href
  {https://doi.org/10.1103/PhysRevA.93.043607} {\bibfield  {journal} {\bibinfo
  {journal} {Physical Review A}\ }\textbf {\bibinfo {volume} {93}},\ \bibinfo
  {pages} {043607} (\bibinfo {year} {2016})}\BibitemShut {NoStop}%
\bibitem [{\citenamefont {Bouton}\ \emph {et~al.}(2020)\citenamefont {Bouton},
  \citenamefont {Nettersheim}, \citenamefont {Adam}, \citenamefont {Schmidt},
  \citenamefont {Mayer}, \citenamefont {Lausch}, \citenamefont {Tiemann},\ and\
  \citenamefont {Widera}}]{Bouton2020}%
  \BibitemOpen
  \bibfield  {author} {\bibinfo {author} {\bibfnamefont {Q.}~\bibnamefont
  {Bouton}}, \bibinfo {author} {\bibfnamefont {J.}~\bibnamefont {Nettersheim}},
  \bibinfo {author} {\bibfnamefont {D.}~\bibnamefont {Adam}}, \bibinfo {author}
  {\bibfnamefont {F.}~\bibnamefont {Schmidt}}, \bibinfo {author} {\bibfnamefont
  {D.}~\bibnamefont {Mayer}}, \bibinfo {author} {\bibfnamefont
  {T.}~\bibnamefont {Lausch}}, \bibinfo {author} {\bibfnamefont
  {E.}~\bibnamefont {Tiemann}},\ and\ \bibinfo {author} {\bibfnamefont
  {A.}~\bibnamefont {Widera}},\ }\bibfield  {title} {\bibinfo {title}
  {Single-atom quantum probes for ultracold gases boosted by nonequilibrium
  spin dynamics},\ }\href {https://doi.org/10.1103/PhysRevX.10.011018}
  {\bibfield  {journal} {\bibinfo  {journal} {Physical Review X}\ }\textbf
  {\bibinfo {volume} {10}},\ \bibinfo {pages} {011018} (\bibinfo {year}
  {2020})}\BibitemShut {NoStop}%
\bibitem [{\citenamefont {Adam}\ \emph {et~al.}(2022)\citenamefont {Adam},
  \citenamefont {Bouton}, \citenamefont {Nettersheim}, \citenamefont
  {Burgardt},\ and\ \citenamefont {Widera}}]{Adam2022}%
  \BibitemOpen
  \bibfield  {author} {\bibinfo {author} {\bibfnamefont {D.}~\bibnamefont
  {Adam}}, \bibinfo {author} {\bibfnamefont {Q.}~\bibnamefont {Bouton}},
  \bibinfo {author} {\bibfnamefont {J.}~\bibnamefont {Nettersheim}}, \bibinfo
  {author} {\bibfnamefont {S.}~\bibnamefont {Burgardt}},\ and\ \bibinfo
  {author} {\bibfnamefont {A.}~\bibnamefont {Widera}},\ }\bibfield  {title}
  {\bibinfo {title} {Coherent and dephasing spectroscopy for single-impurity
  probing of an ultracold bath},\ }\href
  {https://doi.org/10.1103/PhysRevLett.129.120404} {\bibfield  {journal}
  {\bibinfo  {journal} {Physical Review Letters}\ }\textbf {\bibinfo {volume}
  {129}},\ \bibinfo {pages} {120404} (\bibinfo {year} {2022})}\BibitemShut
  {NoStop}%
\bibitem [{\citenamefont {Hovhannisyan}\ and\ \citenamefont
  {Correa}(2018)}]{Hovhannisyan2018}%
  \BibitemOpen
  \bibfield  {author} {\bibinfo {author} {\bibfnamefont {K.~V.}\ \bibnamefont
  {Hovhannisyan}}\ and\ \bibinfo {author} {\bibfnamefont {L.~A.}\ \bibnamefont
  {Correa}},\ }\bibfield  {title} {\bibinfo {title} {Measuring the temperature
  of cold many-body quantum systems},\ }\href
  {https://doi.org/10.1103/PhysRevB.98.045101} {\bibfield  {journal} {\bibinfo
  {journal} {Physical Review B}\ }\textbf {\bibinfo {volume} {98}},\ \bibinfo
  {pages} {045101} (\bibinfo {year} {2018})}\BibitemShut {NoStop}%
\bibitem [{\citenamefont {Potts}\ \emph {et~al.}(2019)\citenamefont {Potts},
  \citenamefont {Brask},\ and\ \citenamefont {Brunner}}]{Potts2019}%
  \BibitemOpen
  \bibfield  {author} {\bibinfo {author} {\bibfnamefont {P.~P.}\ \bibnamefont
  {Potts}}, \bibinfo {author} {\bibfnamefont {J.~B.}\ \bibnamefont {Brask}},\
  and\ \bibinfo {author} {\bibfnamefont {N.}~\bibnamefont {Brunner}},\
  }\bibfield  {title} {\bibinfo {title} {Fundamental limits on low-temperature
  quantum thermometry with finite resolution},\ }\href
  {https://doi.org/10.22331/q-2019-07-09-161} {\bibfield  {journal} {\bibinfo
  {journal} {Quantum}\ }\textbf {\bibinfo {volume} {3}},\ \bibinfo {pages}
  {161} (\bibinfo {year} {2019})}\BibitemShut {NoStop}%
\bibitem [{\citenamefont {J{\o}rgensen}\ \emph {et~al.}(2020)\citenamefont
  {J{\o}rgensen}, \citenamefont {Potts}, \citenamefont {Paris},\ and\
  \citenamefont {Brask}}]{Jorgensen2020}%
  \BibitemOpen
  \bibfield  {author} {\bibinfo {author} {\bibfnamefont {M.~R.}\ \bibnamefont
  {J{\o}rgensen}}, \bibinfo {author} {\bibfnamefont {P.~P.}\ \bibnamefont
  {Potts}}, \bibinfo {author} {\bibfnamefont {M.~G.~A.}\ \bibnamefont
  {Paris}},\ and\ \bibinfo {author} {\bibfnamefont {J.~B.}\ \bibnamefont
  {Brask}},\ }\bibfield  {title} {\bibinfo {title} {Tight bound on
  finite-resolution quantum thermometry at low temperatures},\ }\href
  {https://doi.org/10.1103/PhysRevResearch.2.033394} {\bibfield  {journal}
  {\bibinfo  {journal} {Physical Review Research}\ }\textbf {\bibinfo {volume}
  {2}},\ \bibinfo {pages} {033394} (\bibinfo {year} {2020})}\BibitemShut
  {NoStop}%
\bibitem [{\citenamefont {Mukherjee}\ \emph {et~al.}(2019)\citenamefont
  {Mukherjee}, \citenamefont {Zwick}, \citenamefont {Ghosh}, \citenamefont
  {Chen},\ and\ \citenamefont {Kurizki}}]{Mukherjee2019}%
  \BibitemOpen
  \bibfield  {author} {\bibinfo {author} {\bibfnamefont {V.}~\bibnamefont
  {Mukherjee}}, \bibinfo {author} {\bibfnamefont {A.}~\bibnamefont {Zwick}},
  \bibinfo {author} {\bibfnamefont {A.}~\bibnamefont {Ghosh}}, \bibinfo
  {author} {\bibfnamefont {X.}~\bibnamefont {Chen}},\ and\ \bibinfo {author}
  {\bibfnamefont {G.}~\bibnamefont {Kurizki}},\ }\bibfield  {title} {\bibinfo
  {title} {Enhanced precision bound of low-temperature quantum thermometry via
  dynamical control},\ }\href {https://doi.org/10.1038/s42005-019-0265-y}
  {\bibfield  {journal} {\bibinfo  {journal} {Communications Physics}\ }\textbf
  {\bibinfo {volume} {2}},\ \bibinfo {pages} {162} (\bibinfo {year}
  {2019})}\BibitemShut {NoStop}%
\bibitem [{\citenamefont {Glatthard}\ and\ \citenamefont
  {Correa}(2022)}]{Glatthard2022}%
  \BibitemOpen
  \bibfield  {author} {\bibinfo {author} {\bibfnamefont {J.}~\bibnamefont
  {Glatthard}}\ and\ \bibinfo {author} {\bibfnamefont {L.~A.}\ \bibnamefont
  {Correa}},\ }\bibfield  {title} {\bibinfo {title} {Bending the rules of
  low-temperature thermometry with periodic driving},\ }\href
  {https://doi.org/10.22331/q-2022-05-09-705} {\bibfield  {journal} {\bibinfo
  {journal} {Quantum}\ }\textbf {\bibinfo {volume} {6}},\ \bibinfo {pages}
  {705} (\bibinfo {year} {2022})}\BibitemShut {NoStop}%
\bibitem [{\citenamefont {Correa}\ \emph {et~al.}(2017)\citenamefont {Correa},
  \citenamefont {Perarnau-Llobet}, \citenamefont {Hovhannisyan}, \citenamefont
  {Hern{\'a}ndez-Santana}, \citenamefont {Mehboudi},\ and\ \citenamefont
  {Sanpera}}]{Correa2017}%
  \BibitemOpen
  \bibfield  {author} {\bibinfo {author} {\bibfnamefont {L.~A.}\ \bibnamefont
  {Correa}}, \bibinfo {author} {\bibfnamefont {M.}~\bibnamefont
  {Perarnau-Llobet}}, \bibinfo {author} {\bibfnamefont {K.~V.}\ \bibnamefont
  {Hovhannisyan}}, \bibinfo {author} {\bibfnamefont {S.}~\bibnamefont
  {Hern{\'a}ndez-Santana}}, \bibinfo {author} {\bibfnamefont {M.}~\bibnamefont
  {Mehboudi}},\ and\ \bibinfo {author} {\bibfnamefont {A.}~\bibnamefont
  {Sanpera}},\ }\bibfield  {title} {\bibinfo {title} {Enhancement of
  low-temperature thermometry by strong coupling},\ }\href
  {https://doi.org/10.1103/PhysRevA.96.062103} {\bibfield  {journal} {\bibinfo
  {journal} {Physical Review A}\ }\textbf {\bibinfo {volume} {96}},\ \bibinfo
  {pages} {062103} (\bibinfo {year} {2017})}\BibitemShut {NoStop}%
\bibitem [{\citenamefont {Mehboudi}\ \emph
  {et~al.}(2019{\natexlab{b}})\citenamefont {Mehboudi}, \citenamefont {Lampo},
  \citenamefont {Charalambous}, \citenamefont {Correa}, \citenamefont
  {Garc{\'i}a-March},\ and\ \citenamefont {Lewenstein}}]{Mehboudi2019Polarons}%
  \BibitemOpen
  \bibfield  {author} {\bibinfo {author} {\bibfnamefont {M.}~\bibnamefont
  {Mehboudi}}, \bibinfo {author} {\bibfnamefont {A.}~\bibnamefont {Lampo}},
  \bibinfo {author} {\bibfnamefont {C.}~\bibnamefont {Charalambous}}, \bibinfo
  {author} {\bibfnamefont {L.~A.}\ \bibnamefont {Correa}}, \bibinfo {author}
  {\bibfnamefont {M.~A.}\ \bibnamefont {Garc{\'i}a-March}},\ and\ \bibinfo
  {author} {\bibfnamefont {M.}~\bibnamefont {Lewenstein}},\ }\bibfield  {title}
  {\bibinfo {title} {Using polarons for sub-nk quantum nondemolition
  thermometry in a bose-einstein condensate},\ }\href
  {https://doi.org/10.1103/PhysRevLett.122.030403} {\bibfield  {journal}
  {\bibinfo  {journal} {Physical Review Letters}\ }\textbf {\bibinfo {volume}
  {122}},\ \bibinfo {pages} {030403} (\bibinfo {year}
  {2019}{\natexlab{b}})}\BibitemShut {NoStop}%
\bibitem [{\citenamefont {Planella}\ \emph {et~al.}(2022)\citenamefont
  {Planella}, \citenamefont {Cenni}, \citenamefont {Ac{\'i}n},\ and\
  \citenamefont {Mehboudi}}]{Planella2022}%
  \BibitemOpen
  \bibfield  {author} {\bibinfo {author} {\bibfnamefont {G.}~\bibnamefont
  {Planella}}, \bibinfo {author} {\bibfnamefont {M.~F.~B.}\ \bibnamefont
  {Cenni}}, \bibinfo {author} {\bibfnamefont {A.}~\bibnamefont {Ac{\'i}n}},\
  and\ \bibinfo {author} {\bibfnamefont {M.}~\bibnamefont {Mehboudi}},\
  }\bibfield  {title} {\bibinfo {title} {Bath-induced correlations enhance
  thermometry precision at low temperatures},\ }\href
  {https://doi.org/10.1103/PhysRevLett.128.040502} {\bibfield  {journal}
  {\bibinfo  {journal} {Physical Review Letters}\ }\textbf {\bibinfo {volume}
  {128}},\ \bibinfo {pages} {040502} (\bibinfo {year} {2022})}\BibitemShut
  {NoStop}%
\bibitem [{\citenamefont {Khan}\ \emph {et~al.}(2022)\citenamefont {Khan},
  \citenamefont {Mehboudi}, \citenamefont {Ter{\c{c}}as}, \citenamefont
  {Lewenstein},\ and\ \citenamefont {Garc{\'i}a-March}}]{Khan2022}%
  \BibitemOpen
  \bibfield  {author} {\bibinfo {author} {\bibfnamefont {M.~M.}\ \bibnamefont
  {Khan}}, \bibinfo {author} {\bibfnamefont {M.}~\bibnamefont {Mehboudi}},
  \bibinfo {author} {\bibfnamefont {H.}~\bibnamefont {Ter{\c{c}}as}}, \bibinfo
  {author} {\bibfnamefont {M.}~\bibnamefont {Lewenstein}},\ and\ \bibinfo
  {author} {\bibfnamefont {M.~A.}\ \bibnamefont {Garc{\'i}a-March}},\
  }\bibfield  {title} {\bibinfo {title} {Subnanokelvin thermometry of an
  interacting $d$-dimensional homogeneous bose gas},\ }\href
  {https://doi.org/10.1103/PhysRevResearch.4.023191} {\bibfield  {journal}
  {\bibinfo  {journal} {Physical Review Research}\ }\textbf {\bibinfo {volume}
  {4}},\ \bibinfo {pages} {023191} (\bibinfo {year} {2022})}\BibitemShut
  {NoStop}%
\bibitem [{\citenamefont {Brunelli}\ \emph {et~al.}(2011)\citenamefont
  {Brunelli}, \citenamefont {Olivares},\ and\ \citenamefont
  {Paris}}]{Brunelli2011}%
  \BibitemOpen
  \bibfield  {author} {\bibinfo {author} {\bibfnamefont {M.}~\bibnamefont
  {Brunelli}}, \bibinfo {author} {\bibfnamefont {S.}~\bibnamefont {Olivares}},\
  and\ \bibinfo {author} {\bibfnamefont {M.~G.~A.}\ \bibnamefont {Paris}},\
  }\bibfield  {title} {\bibinfo {title} {Qubit thermometry for micromechanical
  resonators},\ }\href {https://doi.org/10.1103/PhysRevA.84.032105} {\bibfield
  {journal} {\bibinfo  {journal} {Physical Review A}\ }\textbf {\bibinfo
  {volume} {84}},\ \bibinfo {pages} {032105} (\bibinfo {year}
  {2011})}\BibitemShut {NoStop}%
\bibitem [{\citenamefont {Brunelli}\ \emph {et~al.}(2012)\citenamefont
  {Brunelli}, \citenamefont {Olivares}, \citenamefont {Paternostro},\ and\
  \citenamefont {Paris}}]{Brunelli2012}%
  \BibitemOpen
  \bibfield  {author} {\bibinfo {author} {\bibfnamefont {M.}~\bibnamefont
  {Brunelli}}, \bibinfo {author} {\bibfnamefont {S.}~\bibnamefont {Olivares}},
  \bibinfo {author} {\bibfnamefont {M.}~\bibnamefont {Paternostro}},\ and\
  \bibinfo {author} {\bibfnamefont {M.~G.~A.}\ \bibnamefont {Paris}},\
  }\bibfield  {title} {\bibinfo {title} {Qubit-assisted thermometry of a
  quantum harmonic oscillator},\ }\href
  {https://doi.org/10.1103/PhysRevA.86.012125} {\bibfield  {journal} {\bibinfo
  {journal} {Physical Review A}\ }\textbf {\bibinfo {volume} {86}},\ \bibinfo
  {pages} {012125} (\bibinfo {year} {2012})}\BibitemShut {NoStop}%
\bibitem [{\citenamefont {Jevtic}\ \emph {et~al.}(2015)\citenamefont {Jevtic},
  \citenamefont {Newman}, \citenamefont {Rudolph},\ and\ \citenamefont
  {Stace}}]{Jevtic2015}%
  \BibitemOpen
  \bibfield  {author} {\bibinfo {author} {\bibfnamefont {S.}~\bibnamefont
  {Jevtic}}, \bibinfo {author} {\bibfnamefont {D.}~\bibnamefont {Newman}},
  \bibinfo {author} {\bibfnamefont {T.}~\bibnamefont {Rudolph}},\ and\ \bibinfo
  {author} {\bibfnamefont {T.~M.}\ \bibnamefont {Stace}},\ }\bibfield  {title}
  {\bibinfo {title} {Single-qubit thermometry},\ }\href
  {https://doi.org/10.1103/PhysRevA.91.012331} {\bibfield  {journal} {\bibinfo
  {journal} {Physical Review A}\ }\textbf {\bibinfo {volume} {91}},\ \bibinfo
  {pages} {012331} (\bibinfo {year} {2015})}\BibitemShut {NoStop}%
\bibitem [{\citenamefont {Guo}\ \emph {et~al.}(2015)\citenamefont {Guo},
  \citenamefont {Xu}, \citenamefont {Zou},\ and\ \citenamefont
  {Shao}}]{Guo2015}%
  \BibitemOpen
  \bibfield  {author} {\bibinfo {author} {\bibfnamefont {L.-S.}\ \bibnamefont
  {Guo}}, \bibinfo {author} {\bibfnamefont {B.-M.}\ \bibnamefont {Xu}},
  \bibinfo {author} {\bibfnamefont {J.}~\bibnamefont {Zou}},\ and\ \bibinfo
  {author} {\bibfnamefont {B.}~\bibnamefont {Shao}},\ }\bibfield  {title}
  {\bibinfo {title} {Improved thermometry of low-temperature quantum systems by
  a ring-structure probe},\ }\href {https://doi.org/10.1103/PhysRevA.92.052112}
  {\bibfield  {journal} {\bibinfo  {journal} {Physical Review A}\ }\textbf
  {\bibinfo {volume} {92}},\ \bibinfo {pages} {052112} (\bibinfo {year}
  {2015})}\BibitemShut {NoStop}%
\bibitem [{\citenamefont {De~Pasquale}\ \emph {et~al.}(2017)\citenamefont
  {De~Pasquale}, \citenamefont {Yuasa},\ and\ \citenamefont
  {Giovannetti}}]{DePasquale2017}%
  \BibitemOpen
  \bibfield  {author} {\bibinfo {author} {\bibfnamefont {A.}~\bibnamefont
  {De~Pasquale}}, \bibinfo {author} {\bibfnamefont {K.}~\bibnamefont {Yuasa}},\
  and\ \bibinfo {author} {\bibfnamefont {V.}~\bibnamefont {Giovannetti}},\
  }\bibfield  {title} {\bibinfo {title} {Estimating temperature via sequential
  measurements},\ }\href {https://doi.org/10.1103/PhysRevA.96.012316}
  {\bibfield  {journal} {\bibinfo  {journal} {Physical Review A}\ }\textbf
  {\bibinfo {volume} {96}},\ \bibinfo {pages} {012316} (\bibinfo {year}
  {2017})}\BibitemShut {NoStop}%
\bibitem [{\citenamefont {Hofer}\ \emph {et~al.}(2017)\citenamefont {Hofer},
  \citenamefont {Brask}, \citenamefont {Perarnau-Llobet},\ and\ \citenamefont
  {Brunner}}]{Hofer2017}%
  \BibitemOpen
  \bibfield  {author} {\bibinfo {author} {\bibfnamefont {P.~P.}\ \bibnamefont
  {Hofer}}, \bibinfo {author} {\bibfnamefont {J.~B.}\ \bibnamefont {Brask}},
  \bibinfo {author} {\bibfnamefont {M.}~\bibnamefont {Perarnau-Llobet}},\ and\
  \bibinfo {author} {\bibfnamefont {N.}~\bibnamefont {Brunner}},\ }\bibfield
  {title} {\bibinfo {title} {Quantum thermal machine as a thermometer},\ }\href
  {https://doi.org/10.1103/PhysRevLett.119.090603} {\bibfield  {journal}
  {\bibinfo  {journal} {Physical Review Letters}\ }\textbf {\bibinfo {volume}
  {119}},\ \bibinfo {pages} {090603} (\bibinfo {year} {2017})}\BibitemShut
  {NoStop}%
\bibitem [{\citenamefont {Cavina}\ \emph {et~al.}(2018)\citenamefont {Cavina},
  \citenamefont {Mancino}, \citenamefont {De~Pasquale}, \citenamefont
  {Gianani}, \citenamefont {Sbroscia}, \citenamefont {Booth}, \citenamefont
  {Roccia}, \citenamefont {Raimondi}, \citenamefont {Giovannetti},\ and\
  \citenamefont {Barbieri}}]{Cavina2018}%
  \BibitemOpen
  \bibfield  {author} {\bibinfo {author} {\bibfnamefont {V.}~\bibnamefont
  {Cavina}}, \bibinfo {author} {\bibfnamefont {L.}~\bibnamefont {Mancino}},
  \bibinfo {author} {\bibfnamefont {A.}~\bibnamefont {De~Pasquale}}, \bibinfo
  {author} {\bibfnamefont {I.}~\bibnamefont {Gianani}}, \bibinfo {author}
  {\bibfnamefont {M.}~\bibnamefont {Sbroscia}}, \bibinfo {author}
  {\bibfnamefont {R.~I.}\ \bibnamefont {Booth}}, \bibinfo {author}
  {\bibfnamefont {E.}~\bibnamefont {Roccia}}, \bibinfo {author} {\bibfnamefont
  {R.}~\bibnamefont {Raimondi}}, \bibinfo {author} {\bibfnamefont
  {V.}~\bibnamefont {Giovannetti}},\ and\ \bibinfo {author} {\bibfnamefont
  {M.}~\bibnamefont {Barbieri}},\ }\bibfield  {title} {\bibinfo {title}
  {Bridging thermodynamics and metrology in nonequilibrium quantum
  thermometry},\ }\href {https://doi.org/10.1103/PhysRevA.98.050101} {\bibfield
   {journal} {\bibinfo  {journal} {Physical Review A}\ }\textbf {\bibinfo
  {volume} {98}},\ \bibinfo {pages} {050101} (\bibinfo {year}
  {2018})}\BibitemShut {NoStop}%
\bibitem [{\citenamefont {Mancino}\ \emph {et~al.}(2020)\citenamefont
  {Mancino}, \citenamefont {Genoni}, \citenamefont {Barbieri},\ and\
  \citenamefont {Paternostro}}]{Mancino2020}%
  \BibitemOpen
  \bibfield  {author} {\bibinfo {author} {\bibfnamefont {L.}~\bibnamefont
  {Mancino}}, \bibinfo {author} {\bibfnamefont {M.~G.}\ \bibnamefont {Genoni}},
  \bibinfo {author} {\bibfnamefont {M.}~\bibnamefont {Barbieri}},\ and\
  \bibinfo {author} {\bibfnamefont {M.}~\bibnamefont {Paternostro}},\
  }\bibfield  {title} {\bibinfo {title} {Nonequilibrium readiness and precision
  of gaussian quantum thermometers},\ }\href
  {https://doi.org/10.1103/PhysRevResearch.2.033498} {\bibfield  {journal}
  {\bibinfo  {journal} {Physical Review Research}\ }\textbf {\bibinfo {volume}
  {2}},\ \bibinfo {pages} {033498} (\bibinfo {year} {2020})}\BibitemShut
  {NoStop}%
\bibitem [{\citenamefont {Mitchison}\ \emph {et~al.}(2020)\citenamefont
  {Mitchison}, \citenamefont {Fogarty}, \citenamefont {Guarnieri},
  \citenamefont {Campbell}, \citenamefont {Busch},\ and\ \citenamefont
  {Goold}}]{Mitchison2020}%
  \BibitemOpen
  \bibfield  {author} {\bibinfo {author} {\bibfnamefont {M.~T.}\ \bibnamefont
  {Mitchison}}, \bibinfo {author} {\bibfnamefont {T.}~\bibnamefont {Fogarty}},
  \bibinfo {author} {\bibfnamefont {G.}~\bibnamefont {Guarnieri}}, \bibinfo
  {author} {\bibfnamefont {S.}~\bibnamefont {Campbell}}, \bibinfo {author}
  {\bibfnamefont {T.}~\bibnamefont {Busch}},\ and\ \bibinfo {author}
  {\bibfnamefont {J.}~\bibnamefont {Goold}},\ }\bibfield  {title} {\bibinfo
  {title} {In situ thermometry of a cold fermi gas via dephasing impurities},\
  }\href {https://doi.org/10.1103/PhysRevLett.125.080402} {\bibfield  {journal}
  {\bibinfo  {journal} {Physical Review Letters}\ }\textbf {\bibinfo {volume}
  {125}},\ \bibinfo {pages} {080402} (\bibinfo {year} {2020})}\BibitemShut
  {NoStop}%
\bibitem [{\citenamefont {Hovhannisyan}\ \emph {et~al.}(2021)\citenamefont
  {Hovhannisyan}, \citenamefont {J{\o}rgensen}, \citenamefont {Landi},
  \citenamefont {Alhambra}, \citenamefont {Brask},\ and\ \citenamefont
  {Perarnau-Llobet}}]{Hovhannisyan2021}%
  \BibitemOpen
  \bibfield  {author} {\bibinfo {author} {\bibfnamefont {K.~V.}\ \bibnamefont
  {Hovhannisyan}}, \bibinfo {author} {\bibfnamefont {M.~R.}\ \bibnamefont
  {J{\o}rgensen}}, \bibinfo {author} {\bibfnamefont {G.~T.}\ \bibnamefont
  {Landi}}, \bibinfo {author} {\bibfnamefont {{\'A}.~M.}\ \bibnamefont
  {Alhambra}}, \bibinfo {author} {\bibfnamefont {J.~B.}\ \bibnamefont
  {Brask}},\ and\ \bibinfo {author} {\bibfnamefont {M.}~\bibnamefont
  {Perarnau-Llobet}},\ }\bibfield  {title} {\bibinfo {title} {Optimal quantum
  thermometry with coarse-grained measurements},\ }\href
  {https://doi.org/10.1103/PRXQuantum.2.020322} {\bibfield  {journal} {\bibinfo
   {journal} {PRX Quantum}\ }\textbf {\bibinfo {volume} {2}},\ \bibinfo {pages}
  {020322} (\bibinfo {year} {2021})}\BibitemShut {NoStop}%
\bibitem [{\citenamefont {Zhang}\ and\ \citenamefont {Tong}(2022)}]{Zhang2022}%
  \BibitemOpen
  \bibfield  {author} {\bibinfo {author} {\bibfnamefont {D.-J.}\ \bibnamefont
  {Zhang}}\ and\ \bibinfo {author} {\bibfnamefont {D.~M.}\ \bibnamefont
  {Tong}},\ }\bibfield  {title} {\bibinfo {title} {Approaching
  heisenberg-scalable thermometry with built-in robustness against noise},\
  }\href {https://doi.org/10.1038/s41534-022-00588-2} {\bibfield  {journal}
  {\bibinfo  {journal} {npj Quantum Information}\ }\textbf {\bibinfo {volume}
  {8}},\ \bibinfo {pages} {81} (\bibinfo {year} {2022})}\BibitemShut {NoStop}%
\bibitem [{\citenamefont {Correa}\ \emph {et~al.}(2015)\citenamefont {Correa},
  \citenamefont {Mehboudi}, \citenamefont {Adesso},\ and\ \citenamefont
  {Sanpera}}]{Correa2015}%
  \BibitemOpen
  \bibfield  {author} {\bibinfo {author} {\bibfnamefont {L.~A.}\ \bibnamefont
  {Correa}}, \bibinfo {author} {\bibfnamefont {M.}~\bibnamefont {Mehboudi}},
  \bibinfo {author} {\bibfnamefont {G.}~\bibnamefont {Adesso}},\ and\ \bibinfo
  {author} {\bibfnamefont {A.}~\bibnamefont {Sanpera}},\ }\bibfield  {title}
  {\bibinfo {title} {Individual quantum probes for optimal thermometry},\
  }\href {https://doi.org/10.1103/PhysRevLett.114.220405} {\bibfield  {journal}
  {\bibinfo  {journal} {Phys. Rev. Lett.}\ }\textbf {\bibinfo {volume} {114}},\
  \bibinfo {pages} {220405} (\bibinfo {year} {2015})}\BibitemShut {NoStop}%
\bibitem [{\citenamefont {Campbell}\ \emph {et~al.}(2018)\citenamefont
  {Campbell}, \citenamefont {Genoni},\ and\ \citenamefont
  {Deffner}}]{Campbell2018}%
  \BibitemOpen
  \bibfield  {author} {\bibinfo {author} {\bibfnamefont {S.}~\bibnamefont
  {Campbell}}, \bibinfo {author} {\bibfnamefont {M.~G.}\ \bibnamefont
  {Genoni}},\ and\ \bibinfo {author} {\bibfnamefont {S.}~\bibnamefont
  {Deffner}},\ }\bibfield  {title} {\bibinfo {title} {Precision thermometry and
  the quantum speed limit},\ }\href {https://doi.org/10.1088/2058-9565/aa9ef4}
  {\bibfield  {journal} {\bibinfo  {journal} {Quantum Science and Technology}\
  }\textbf {\bibinfo {volume} {3}},\ \bibinfo {pages} {025002} (\bibinfo {year}
  {2018})}\BibitemShut {NoStop}%
\bibitem [{\citenamefont {Mok}\ \emph {et~al.}(2021)\citenamefont {Mok},
  \citenamefont {Bharti}, \citenamefont {Kwek},\ and\ \citenamefont
  {Bayat}}]{Mok2021}%
  \BibitemOpen
  \bibfield  {author} {\bibinfo {author} {\bibfnamefont {W.-K.}\ \bibnamefont
  {Mok}}, \bibinfo {author} {\bibfnamefont {K.}~\bibnamefont {Bharti}},
  \bibinfo {author} {\bibfnamefont {L.-C.}\ \bibnamefont {Kwek}},\ and\
  \bibinfo {author} {\bibfnamefont {A.}~\bibnamefont {Bayat}},\ }\bibfield
  {title} {\bibinfo {title} {Optimal probes for global quantum thermometry},\
  }\href {https://doi.org/10.1038/s42005-021-00572-w} {\bibfield  {journal}
  {\bibinfo  {journal} {Communications Physics}\ }\textbf {\bibinfo {volume}
  {4}},\ \bibinfo {pages} {62} (\bibinfo {year} {2021})}\BibitemShut {NoStop}%
\bibitem [{\citenamefont {Sekatski}\ and\ \citenamefont
  {Perarnau-Llobet}(2022)}]{Sekatski2022}%
  \BibitemOpen
  \bibfield  {author} {\bibinfo {author} {\bibfnamefont {P.}~\bibnamefont
  {Sekatski}}\ and\ \bibinfo {author} {\bibfnamefont {M.}~\bibnamefont
  {Perarnau-Llobet}},\ }\bibfield  {title} {\bibinfo {title} {Optimal
  nonequilibrium thermometry in markovian environments},\ }\href
  {https://doi.org/10.22331/q-2022-12-07-869} {\bibfield  {journal} {\bibinfo
  {journal} {Quantum}\ }\textbf {\bibinfo {volume} {6}},\ \bibinfo {pages}
  {869} (\bibinfo {year} {2022})}\BibitemShut {NoStop}%
\bibitem [{\citenamefont {Mehboudi}\ \emph {et~al.}(2022)\citenamefont
  {Mehboudi}, \citenamefont {J{\o}rgensen}, \citenamefont {Seah}, \citenamefont
  {Brask}, \citenamefont {Ko{\l}ody{\'n}ski},\ and\ \citenamefont
  {Perarnau-Llobet}}]{Mehboudi2022}%
  \BibitemOpen
  \bibfield  {author} {\bibinfo {author} {\bibfnamefont {M.}~\bibnamefont
  {Mehboudi}}, \bibinfo {author} {\bibfnamefont {M.~R.}\ \bibnamefont
  {J{\o}rgensen}}, \bibinfo {author} {\bibfnamefont {S.}~\bibnamefont {Seah}},
  \bibinfo {author} {\bibfnamefont {J.~B.}\ \bibnamefont {Brask}}, \bibinfo
  {author} {\bibfnamefont {J.}~\bibnamefont {Ko{\l}ody{\'n}ski}},\ and\
  \bibinfo {author} {\bibfnamefont {M.}~\bibnamefont {Perarnau-Llobet}},\
  }\bibfield  {title} {\bibinfo {title} {Fundamental limits in bayesian
  thermometry and attainability via adaptive strategies},\ }\href
  {https://doi.org/10.1103/PhysRevLett.128.130502} {\bibfield  {journal}
  {\bibinfo  {journal} {Physical Review Letters}\ }\textbf {\bibinfo {volume}
  {128}},\ \bibinfo {pages} {130502} (\bibinfo {year} {2022})}\BibitemShut
  {NoStop}%
\bibitem [{\citenamefont {Abiuso}\ \emph {et~al.}(2024)\citenamefont {Abiuso},
  \citenamefont {Andrea~Erdman}, \citenamefont {Ronen}, \citenamefont {Noé},
  \citenamefont {Haack},\ and\ \citenamefont {Perarnau-Llobet}}]{Abiuso_2024}%
  \BibitemOpen
  \bibfield  {author} {\bibinfo {author} {\bibfnamefont {P.}~\bibnamefont
  {Abiuso}}, \bibinfo {author} {\bibfnamefont {P.}~\bibnamefont
  {Andrea~Erdman}}, \bibinfo {author} {\bibfnamefont {M.}~\bibnamefont
  {Ronen}}, \bibinfo {author} {\bibfnamefont {F.}~\bibnamefont {Noé}},
  \bibinfo {author} {\bibfnamefont {G.}~\bibnamefont {Haack}},\ and\ \bibinfo
  {author} {\bibfnamefont {M.}~\bibnamefont {Perarnau-Llobet}},\ }\bibfield
  {title} {\bibinfo {title} {Optimal thermometers with spin networks},\ }\href
  {https://doi.org/10.1088/2058-9565/ad37d3} {\bibfield  {journal} {\bibinfo
  {journal} {Quantum Science and Technology}\ }\textbf {\bibinfo {volume}
  {9}},\ \bibinfo {pages} {035008} (\bibinfo {year} {2024})}\BibitemShut
  {NoStop}%
\bibitem [{\citenamefont {Cram{\'e}r}(1999)}]{Cramer1999}%
  \BibitemOpen
  \bibfield  {author} {\bibinfo {author} {\bibfnamefont {H.}~\bibnamefont
  {Cram{\'e}r}},\ }\href@noop {} {\emph {\bibinfo {title} {Mathematical Methods
  of Statistics}}},\ \bibinfo {series} {Princeton Mathematical Series},
  Vol.~\bibinfo {volume} {9}\ (\bibinfo  {publisher} {Princeton University
  Press},\ \bibinfo {year} {1999})\BibitemShut {NoStop}%
\bibitem [{\citenamefont {Braunstein}\ and\ \citenamefont
  {Caves}(1994)}]{BraunsteinCaves1994}%
  \BibitemOpen
  \bibfield  {author} {\bibinfo {author} {\bibfnamefont {S.~L.}\ \bibnamefont
  {Braunstein}}\ and\ \bibinfo {author} {\bibfnamefont {C.~M.}\ \bibnamefont
  {Caves}},\ }\bibfield  {title} {\bibinfo {title} {Statistical distance and
  the geometry of quantum states},\ }\href
  {https://doi.org/10.1103/PhysRevLett.72.3439} {\bibfield  {journal} {\bibinfo
   {journal} {Physical Review Letters}\ }\textbf {\bibinfo {volume} {72}},\
  \bibinfo {pages} {3439} (\bibinfo {year} {1994})}\BibitemShut {NoStop}%
\bibitem [{\citenamefont {Srivastava}\ \emph {et~al.}(2025)\citenamefont
  {Srivastava}, \citenamefont {Bhattacharya}, \citenamefont {Lewenstein},\ and\
  \citenamefont {P{\l{}}odzie{\ifmmode \acute{n}\else
  \'{n}\fi{}}}}]{PhysRevA.111.052216}%
  \BibitemOpen
  \bibfield  {author} {\bibinfo {author} {\bibfnamefont {A.~K.}\ \bibnamefont
  {Srivastava}}, \bibinfo {author} {\bibfnamefont {U.}~\bibnamefont
  {Bhattacharya}}, \bibinfo {author} {\bibfnamefont {M.}~\bibnamefont
  {Lewenstein}},\ and\ \bibinfo {author} {\bibfnamefont {M.}~\bibnamefont
  {P{\l{}}odzie{\ifmmode \acute{n}\else \'{n}\fi{}}}},\ }\bibfield  {title}
  {\bibinfo {title} {Topological quantum thermometry},\ }\href
  {https://doi.org/10.1103/PhysRevA.111.052216} {\bibfield  {journal} {\bibinfo
   {journal} {Phys. Rev. A}\ }\textbf {\bibinfo {volume} {111}},\ \bibinfo
  {pages} {052216} (\bibinfo {year} {2025})}\BibitemShut {NoStop}%
\bibitem [{\citenamefont {Kingma}\ and\ \citenamefont {Ba}(2014)}]{adam}%
  \BibitemOpen
  \bibfield  {author} {\bibinfo {author} {\bibfnamefont {D.~P.}\ \bibnamefont
  {Kingma}}\ and\ \bibinfo {author} {\bibfnamefont {J.}~\bibnamefont {Ba}},\
  }\bibfield  {title} {\bibinfo {title} {Adam: A method for stochastic
  optimization},\ }\href {https://arxiv.org/abs/1412.6980} {\bibfield
  {journal} {\bibinfo  {journal} {arXiv preprint arXiv:1412.6980}\ } (\bibinfo
  {year} {2014})}\BibitemShut {NoStop}%
\end{thebibliography}%

\clearpage
\onecolumngrid

\appendix
\section{Numerical optimization using Adam}\label{adam_details}

The numerical optimization of the different models considered in this work presents a nontrivial task, particularly as the number of spins increases. The optimization was performed primarily using the Adam algorithm implemented in the PyTorch library. The numerical implementation follows a similar structure for all models, with the main difference lying in the computation of the energy spectrum. Depending on the model, the energies are obtained either by direct diagonalization of the Hamiltonian or from the corresponding analytical expressions derived in the main text. The optimization procedure was adapted to each model and system size $n$. In particular, for some parameter regimes the optimization landscape exhibits local optima, making convergence to the optimal solution more challenging. Moreover, increasing $n$ generally results in slower and more difficult convergence. To address this issue, the optimization was performed progressively in $n$, using information from smaller system sizes to guide the optimization of larger systems. Several strategies were employed to improve numerical convergence:

\begin{itemize} 

\item \textbf{Increasing the number of optimization iterations} when convergence was particularly slow. \item \textbf{Adapting the initialization of the Hamiltonian parameters.} Depending on the model, the coefficients were either initialized randomly within a prescribed range or initialized using the optimal parameters obtained for $n-1$ spins. 
\item \textbf{Adjusting the Adam hyperparameters during the optimization.} In particular, the learning rate was sometimes decreased after a given number of iterations, allowing for a smoother convergence at later stages. 
\item \textbf{Rescaling the objective function.} In some cases, the heat capacity was normalized by an estimate of its expected optimal value, for instance using the optimum obtained for $n-1$ spins plus a certain increment.
\item \textbf{Shifting the energy spectrum to improve numerical stability.} For some parameter values, the computation of the probabilities $p_i$ entering the heat capacity led to numerical instabilities and \texttt{NaN} values due to very small quantities. Appropriate shifts of the energy spectrum were therefore employed when evaluating the thermal probabilities, exploiting their invariance under a global energy shift. 
\end{itemize} 
Overall, the optimization was carried out progressively and adaptively, building on the cases with smaller $n$, for which convergence was more readily achieved, and addressing the difficulties encountered along the way until convergence was reached in each case. Both the codes and the results obtained are available upon request.

\section{Alternative 3-local generalization of the Star model}\label{alternative_star}

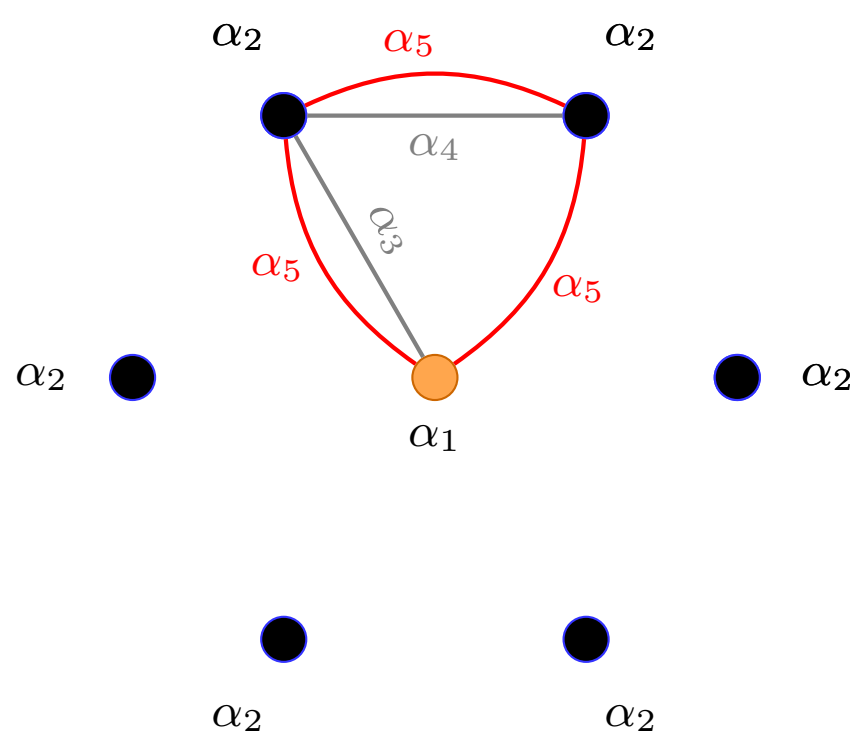
\begin{figure}[h]
\begin{center}
\begin{tikzpicture}
\tikzset{
        nodeA/.style={circle, draw=orange, fill=orange!70, inner sep=3pt},
        nodeB/.style={circle, draw=black, fill=black!80, inner sep=3pt},
        edge/.style={thick, gray},
        triple/.style={thick, red, bend left=20},
    }

% Nodo central (mismo tamaño que los nodos b)
\node[circle, draw=orange!80!black, fill=orange!70, inner sep=3pt] (a) at (0,0) {};
\node at (0,-0.4) {$\alpha_1$}; % etiqueta afuera

% Nodos periféricos y conexiones a 2 cuerpos(gris)
\foreach \i in {0,1,...,8} {
    \node[circle, draw=blue!80, fill=black, inner sep=3pt] (b\i) at (60*\i:2cm) {};
    \node at ($(b\i) + (60*\i:0.6cm)$) {$\alpha_2$};
}
\draw[thick, gray] (a) -- node[midway, above, sloped] {$\alpha_3$} (b2);
\draw[thick, gray] (b1) -- node[midway, below, sloped] {$\alpha_4$} (b2);

% Conexiones a 3 cuerpos (rojo)
\draw[triple] (a) to[bend right=25] node[pos=0.4, right] {$\alpha_5$} (b1);
\draw[triple] (b1) to[bend right=25] node[pos=0.6, above] {$\alpha_5$} (b2);
\draw[triple] (b2) to[bend right=25] node[pos=0.5, left] {$\alpha_5$} (a);

\end{tikzpicture}
\end{center}
\captionsetup{justification=Justified, singlelinecheck=false}\caption{Schematic representation of an alternative 3-local Star model. Gray lines represent two-body interactions, while red lines represent three-body interactions.}
\label{fig:alternative_3star}
\end{figure}

Alternatively, we consider a different extension of the 2-local Star model to three-body interactions. As in the 2-local Star construction, a single qubit plays a distinguished role and interacts with all the remaining qubits. In addition, three-body interactions are introduced by coupling the distinguished qubit to pairs of neighboring qubits. %The resulting construction can thus be viewed as a hybrid between the 2-local Star model and a one-dimensional Ising-like interaction pattern. 
The most general Hamiltonian within this model reads 

\begin{equation} H_{n,3}^{\widetilde{\mathrm{Star}}} = \alpha_1 Z_1 + \alpha_2 \sum_{i=2}^{n} Z_i + \alpha_3 \sum_{i=2}^{n} Z_1 Z_i + \alpha_4 \sum_{i=2}^{n-1} Z_i Z_{i+1} + \alpha_5 \sum_{i=2}^{n-1} Z_1 Z_i Z_{i+1}, \end{equation} 
where $\alpha_i\in\mathbb{R}$ and qubit $1$ denotes the distinguished qubit. Fig.~\ref{fig:alternative_3star} provides a schematic representation of this interaction structure. For clarity, only a subset of the interactions involving the nondistinguished qubits is shown. As in the other Star constructions, the symmetries of the model may impose relations among the coefficients $\alpha_i$. Indeed, numerical optimization for systems of up to $n=10$ qubits consistently yields $\alpha_2=\alpha_3$ and $\alpha_4=\alpha_5$, thereby reducing the number of independent parameters. Nevertheless, this construction exhibits a lower optimal heat capacity than the 3-local All-to-All and Star models. We therefore do not pursue this model further.

%\clearpage
\section{Analytical derivation of the energy spectrum for the $k$-local All-to-All model}\label{app:energy_derivation}
This appendix provides the detailed analytical derivation of the energy levels for the $k$-local All-to-All model introduced in Section~\ref{sec:k_local_AtA}.
Let us consider an arbitrary computational-basis configuration specified by the number of spin-up states, denoted by $p$ (hence $n-p$ represents the number of spin-down states). We associate the Pauli operators $Z_i = \pm 1$ with the spin-up and spin-down states, respectively.The contribution to the energy from a specific $q$-body interaction term depends on the combinations of those $q$ spins. Consider a subset of $q$ interacting spins, out of which $j$ are in the spin-up state and $q-j$ are in the spin-down state. Since each spin-down state contributes a factor of $-1$ to the product $\prod_{i \in S} Z_i$, the term contributes positively if the number of spin-down states ($q-j$) is even, and negatively if it is odd. This sign factor is compactly expressed as $(-1)^{q-j}$.The number of ways to choose $j$ spin-up particles out of the $p$ available spin-up states, and $q-j$ spin-down particles out of the $n-p$ available spin-down states, is given by the product of binomial coefficients $\binom{p}{j}\binom{n-p}{q-j}$.

Summing over all possible values of $j$ from $0$ to $q$, and over all interaction orders $q$ from $1$ to $k$, the complete energy eigenvalue for a configuration with $p$ spins up is:

\begin{equation}
E_p = \sum_{q=1}^{k} \alpha_q \sum_{j=0}^{q} \binom{p}{j}\binom{n-p}{q-j}(-1)^{q-j},
\end{equation}
where each energy level $E_p$ has a degeneracy of $\binom{n}{p}$.

To simplify the evaluation of these sums, we can employ a generating function approach. Let us define the auxiliary polynomial function of variable $x$:

\begin{equation}
f(x) = (1+x)^p(1-x)^{n-p}.
\end{equation}

Using the binomial theorem, $f(x)$ can be expanded as:

\begin{align}
(1+x)^p &= \sum_{j=0}^p \binom{p}{j} x^j, \\ (1-x)^{n-p} &= \sum_{r=0}^{n-p} \binom{n-p}{r} (-1)^r x^r.
\end{align}

Multiplying these two expansions yields:

\begin{equation}
f(x) = \left( \sum_{j=0}^p \binom{p}{j} x^j \right) \left( \sum_{r=0}^{n-p} \binom{n-p}{r} (-1)^r x^r \right).
\end{equation}

Denoting the coefficient of $x^q$ in $f(x)$ by $[x^q] f(x)$, we extract the term corresponding to degree $q$ by setting $r = q-j$:

\begin{equation}
[x^q] = \sum_{j=0}^q \binom{p}{j} \binom{n-p}{q-j} (-1)^{q-j}.
\end{equation}

Consequently, the energy expression can be rewritten concisely in terms of these coefficients:

\begin{equation}
E_p = \sum_{q=1}^{k} \alpha_q [x^q].
\end{equation}

We can evaluate the sum of the coefficients by inspecting the function $f(x)$ at $x = 1$. Specifically, we have:

\begin{equation}
f(1) = 2^p \cdot 0^{n-p} = \sum_{q=0}^n [x^q].
\end{equation}

This leads to two distinct cases depending on $p$:

\begin{itemize}\item \textbf{Case $p = n$:} $f(1) = 2^n$, which implies:

\begin{equation}\sum_{q=0}^n [x^q] = 2^n.
\end{equation}

\item \textbf{Case $0 \le p < n$:} $f(1) = 0$, which implies:

\begin{equation}\sum_{q=0}^n [x^q] = 0.
\end{equation}
\end{itemize}

Now, let us consider the fully connected limit where $k = n$ and all coupling constants are uniform, $\alpha_q = a$. Noting that the $q=0$ term is given by $[x^0] = \binom{p}{0}\binom{n-p}{0}(-1)^0 = 1$, we obtain for the energy:

\begin{equation}E_p = \sum_{q=1}^n a [x^q] = a \left( \sum_{q=0}^n [x^q] - 1 \right).
\end{equation}

Substituting the results for $\sum_{q=0}^n [x^q]$ from the two cases above, we arrive at the exact analytical spectrum:

\begin{equation}
E_p =\begin{cases}(2^n - 1)a, & \text{if } p = n, \\
-a, & \text{if } 0 \le p < n,\end{cases}
\end{equation}

which recovers the globally optimal spectrum consisting of a single non-degenerate level separated from a $(2^n-1)$-degenerate level.

Additionally, it is worth noting that if the coupling constants are chosen as $\alpha_q = (-a)^q$, an analogous derivation can be performed by evaluating the generating function $f(x)$ at $x = -1$. This yields the exact same energy spectrum, thereby revealing a characteristic symmetry of the model that is also consistently observed in numerical simulations.

\end{document}